\documentclass[twocolumn,amsmath,usenatbib]{aastex631}

\hypersetup{linkcolor=red,citecolor=green,filecolor=cyan,urlcolor=magenta}

\newcommand\aastex{AAS\TeX}

\usepackage{soul}
\received{}
\revised{}
\accepted{}

\submitjournal{ApJS}

\shorttitle{\aastex\ Optical variability of eight FRII-type quasars}
\shortauthors{Ku\'zmicz et al.}

\begin{document}

\title{Optical variability of eight FRII-type quasars with 13-yr photometric light curves}

\correspondingauthor{Agnieszka Ku\'zmicz}
\email{cygnus@oa.uj.edu.pl}

\author[0000-0002-3097-5605]{Agnieszka Ku\'zmicz}
\affil{Astronomical Observatory, Jagiellonian University, ul. Orla 171, 30-244 Krakow, Poland}
\affil{Queen Jadwiga Astronomical Observatory in Rzepiennik Biskupi, 33-163 Rzepiennik Strzy\.zewski, Poland}

\author[0000-0002-2224-6664]{Arti Goyal}
\affiliation{Astronomical Observatory, Jagiellonian University, ul. Orla 171, 30-244 Krakow, Poland}

\author[0000-0003-3609-382X]{Stanis{\l}aw Zola}
\affiliation{Astronomical Observatory, Jagiellonian University, ul. Orla 171, 30-244 Krakow, Poland}

\author[0000-0002-0870-7778]{Marek Jamrozy}
\affiliation{Astronomical Observatory, Jagiellonian University, ul. Orla 171, 30-244 Krakow, Poland}

\author[0000-0001-9587-1615]{Marek Dr\'o\.zd\.z}
\affiliation{Mt. Suhora Observatory, Pedagogical University, ul. Podchorazych 2, 30-084
Krakow, Poland}

\author[0000-0002-6293-9940]{Waldemar Og{\l}oza}
\affiliation{Mt. Suhora Observatory, Pedagogical University, ul. Podchorazych 2, 30-084
Krakow, Poland}

\author[0000-0001-5018-3560]{Micha{\l} Siwak}
\affiliation{Konkoly Observatory, Research Centre for Astronomy and Earth Sciences,
E\"otv\"os Lor\'and Research Network (ELKH), Hungarian Academy of Sciences,
Konkoly-Thege Mikl\'os \'ut 15--17, 1121 Budapest, Hungary}

\author{Daniel E. Reichart}
\affiliation{University of North Carolina at Chapel Hill, Chapel Hill, North Carolina, NC 27599, USA}

\author[0000-0003-3642-5484]{Vladimir V. Kouprianov}
\affiliation{University of North Carolina at Chapel Hill, Chapel Hill, North Carolina, NC 27599, USA}

\author{Daniel B. Caton}
\affiliation{Dark Sky Observatory, Department of Physics and Astronomy, Appalachian
State University, Boone, NC 28608, USA}

\begin{abstract}

We characterize the optical variability properties of eight lobe-dominated radio quasars (QSOs): B2\,0709$+$37, FBQS \,J095206.3$+$235245, PG\,1004$+$130, [HB89]\,1156$+$631, [HB89]\,1425$+$267, [HB89]\,1503$+$691, [HB89]\,1721$+$343, 4C\,$+$74.26, systematically monitored for a duration of 13 years since 2009. The quasars are radio-loud objects with extended radio lobes that indicate their orientation close to the sky plane. Five of the eight QSOs are classified as giant radio quasars. All quasars showed variability during our monitoring, with magnitude variations between 0.3 and 1 mag for the least variable and the most variable QSO, respectively. We performed both structure function (SF) analysis and power spectrum density (PSD) analysis for the variability characterization and search for characteristic timescales and periodicities. As a result of our analysis, we obtained relatively steep SF slopes ($\alpha$ ranging from 0.49 to 0.75) that are consistent with the derived PSD slopes ($\sim$2--3). All the PSDs show a good fit to single power law forms, indicating a red-noise character of variability between $\sim$13 years and weeks timescales. We did not measure reliable characteristic timescales of variability from the SF analysis which indicates that the duration of the gathered data is too short to reveal them. The absence of bends in the PSDs (change of slope from $\geq$1 to $\sim$0) on longer timescales indicates that optical variations are most likely caused by thermal instabilities in the accretion disk.

\end{abstract}


\section{Introduction} \label{sec:intro}

Intense emission and its variability on timescales ranging from decades to minutes on many wavebands is the most ubiquitous property of the active galactic nuclei (AGNs). The power source of these sources is believed to be the inflow of matter from the accretion disk onto the supermassive black hole \citep[SMBH; e.g., for a review, ][]{antonucci1993, urry1995}. As matter falls onto the SMBH, most of the plasma is compressed and heated, resulting in radiation from optical to $UV$ wavebands. A small fraction of AGNs also eject bipolar, relativistic plasma jets that travel up to Mpc distances and produce radiation from radio to TeV $\gamma$-ray energies via nonthermal processes occurring in jets \citep[radio-loud AGNs; ][]{hardcastle2018, blandford2019}. 
The flux variability probes the accretion disk or jet close to the central regions of the SMBH, and hence their light curves should exhibit the physical processes responsible for the variations \citep[][]{mangalam1993}.

The cause of long-term optical variability (decades to days) of AGNs is still a matter of debate (e.g., \citealt{hawkins2002}, \citealt{devries2003},  \citealt{vanden2004}, \citealt{bauer2009}, \citealt{macleod2010}, \citealt{schmidt2010}, \citealt{morganson2014}, \citealt{caplar2017}, \citealt{xin2020}). On the basis of theoretical studies, few mechanisms could explain the observed variability. The first mechanism is related to the accretion disk instabilities (e.g. \citealt{kawaguchi1998}, \citealt{siemiginowska1997}), where occasional flare events or blob formation cause a luminosity variation. In the second mechanism named the starburst model (e.g. \citealt{terlevich1992}, \citealt{aretxaga1997}) the flux variations are associated with the evolution of massive stars and supernovae events. In the third mechanism, the observed flux variability is not connected with some intrinsic AGN behavior, but is a result of gravitational microlensing by massive compact objects (e.g. \citealt{hawkins1993}). Each of these scenarios can be differentiated depending on the chromaticity of variations, observed timescales, amplitude of variability, and the distribution of variability amplitudes over different timescales (structure function or power spectrum analysis). For example, for the variability driven by thermal instabilities in the standard Shakura and Sunyaev accretion disk, one expects a relaxation timescale longer for which the flux variations become uncorrelated \citep[$\sim$ few years for 10$^8$ $M_\odot$ SMBH; ][]{hawkins2002, kawaguchi1998, kelly2009}. On the other hand, the starburst model predicts a flattening of the structure function around 100\,d and a logarithmic slope  $\sim$0.8 \citep[][]{kawaguchi1998} while achromatic variations in the light curves obtained at different wavebands are predicted for microlensing events \citep[][]{hawkins1997}.

Although theoretical predictions are well established for the variability of AGN, resulting from instabilities in the accretion disk \citep[][]{mangalam1993}, it is often not possible to give an unambiguous explanation of the observed flux variations. This is because (1) availability of finite-duration light curves with different sampling intervals obtained for inhomogeneously selected source samples, (2) variability analysis methods susceptible to the gaps in the light curves, (3) contrary description of variability from the application of two (or more) different methods, and (4) existence of more than one variability mechanism, each of which may have a different contribution to the observed variability.

Many studies based on radio-loud quasar samples consisting of observations taken on a few epochs but covering a time baseline of a few years show that the ``typical'' variability amplitude is $\sim$0.2--$\sim$0.4\,mag on yearly timescales \citep[][]{netzer1996, garcia1999}. Using the ``ensemble'' variability approach, a quasar light curve consisting of a few thousand data points was derived that covered a time baseline of 3.5\,yr for which the structure function analysis did not reveal any characteristic time scale \citep[that is, a plateau in the SF curve; ][]{vanden2004, bauer2009}. Moreover, the time baseline of optical fluctuations for quasar sources was extended up to 40--50\,yr for which the structure function shows a monotonic and constant increase of variability amplitude with increasing time lag \citep[][]{devries2005}. On the other hand, power spectrum analysis using stochastic modeling of quasar light curves indicates that variability is explained with a damped random walk model up to the time baseline of a few to 10 years with relaxation timescale $\sim$a few years, leading to a
flattening of the power spectrum toward white noise \citep[][]{kelly2009, macleod2012, caplar2017}.
 
Furthermore, the characteristic parameters of the variability are often found to be correlated with the AGN properties. For example, the optical variability amplitude is anticorrelated with the optical luminosity and Eddington ratio (e.g. \citealt{helfand2001}, \citealt{ai2010}) and that it is correlated with black hole mass (e.g. \citealt{wold2007}; \citealt{wilhite2008}).

Dedicated long-term optical monitoring of radio-loud sources has been carried out for large number of sources consisting mostly of blazars (emission dominated by the relativistic jets; \citealt{hovatta2019}) or core-dominated quasars (sources presenting unresolved radio cores in arcsec scale-resolution GHz-band radio images; \citealt{impey1991}), by many groups \citep[e.g., ][]{bonning2012, jorstad2016, nilsson2018}.
This is the first study where we examine the optical variability of a sample of quasars consisting of a modest number of lobe-dominated radio sources observed in our dedicated monitoring program.

We began monitoring the lobe-dominated quasars with large-scale radio lobes with an aim of better understanding the processes underlying the origin of radio lobe expansion at large distances from the host galaxy. One of the hypotheses that may explain the formation of large-scale radio lobes is that the central engine, which is responsible for generating radio jets, has some specific properties. This hypothesis was examined in a series of papers e.g. \cite{ishwara1999}, \cite{kuzmicz2012,kuzmicz2021}, \cite{kuzmicz2019, kuzmicz2021b} and \cite{dabhade2020}. 
However, finding a link between AGN variability and the radio jet's ejection at large distances from the host galaxy requires long-term monitoring of a large sample of such sources.

The main purpose of this work is to characterize the optical variability properties of a sample of eight lobe-dominated radio quasars using 13\,yr-long $R-$band light curves obtained with a sampling interval of few days to weeks. The observed light curves from high-quality data sets (mean photometric accuracy $\sim$0.01-0.1\,mag) were used to study the variability of the quasars. We quantify the optical variability properties of observed quasars using structure function and power spectral density analyses.

\section{Sample}\label{sec:sample}

\begin{figure*}[ht!] 
\centering 
\includegraphics[height=4cm]{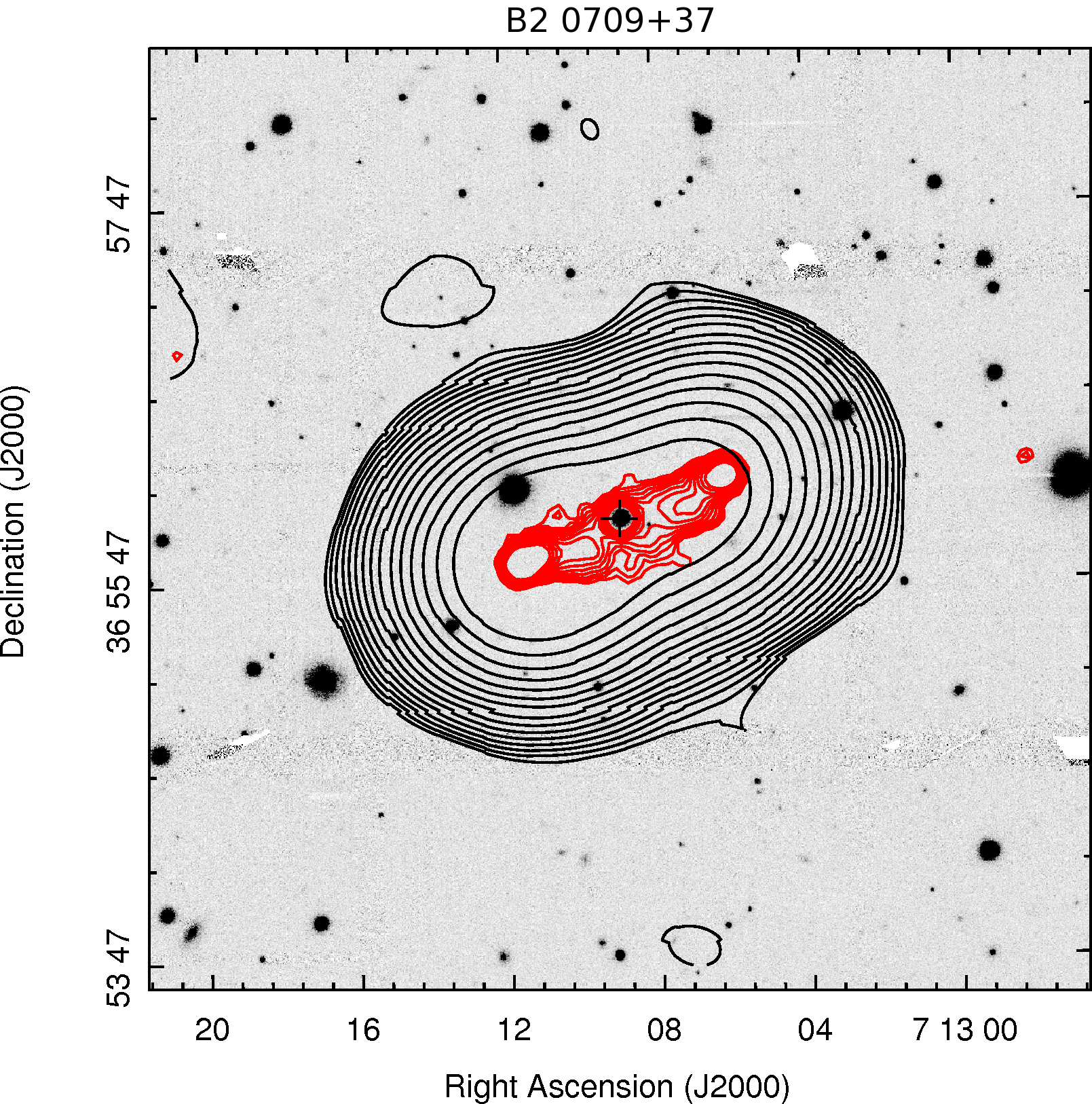}
\includegraphics[height=4cm]{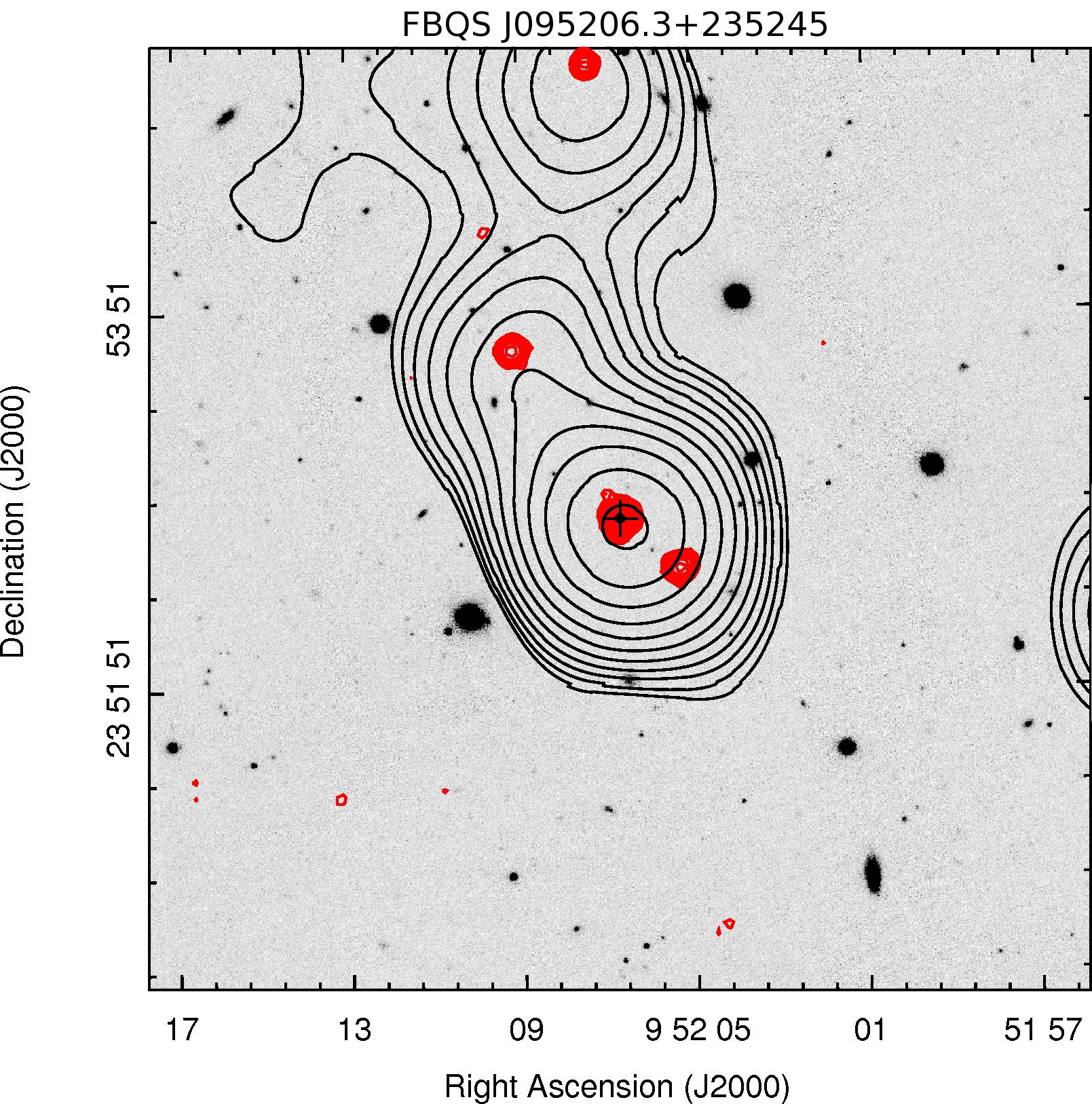}
\includegraphics[height=4cm]{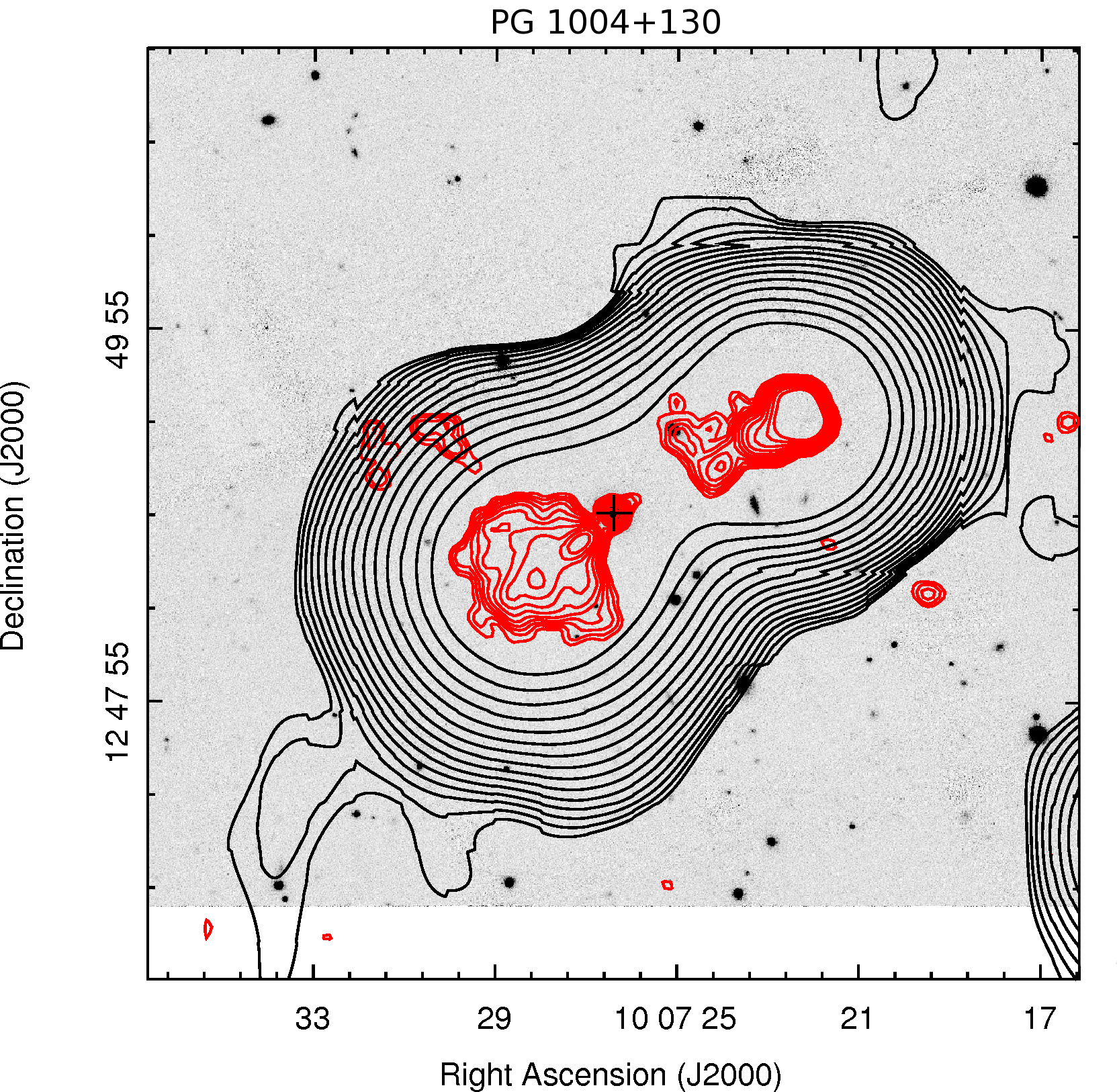}
\includegraphics[height=4cm]{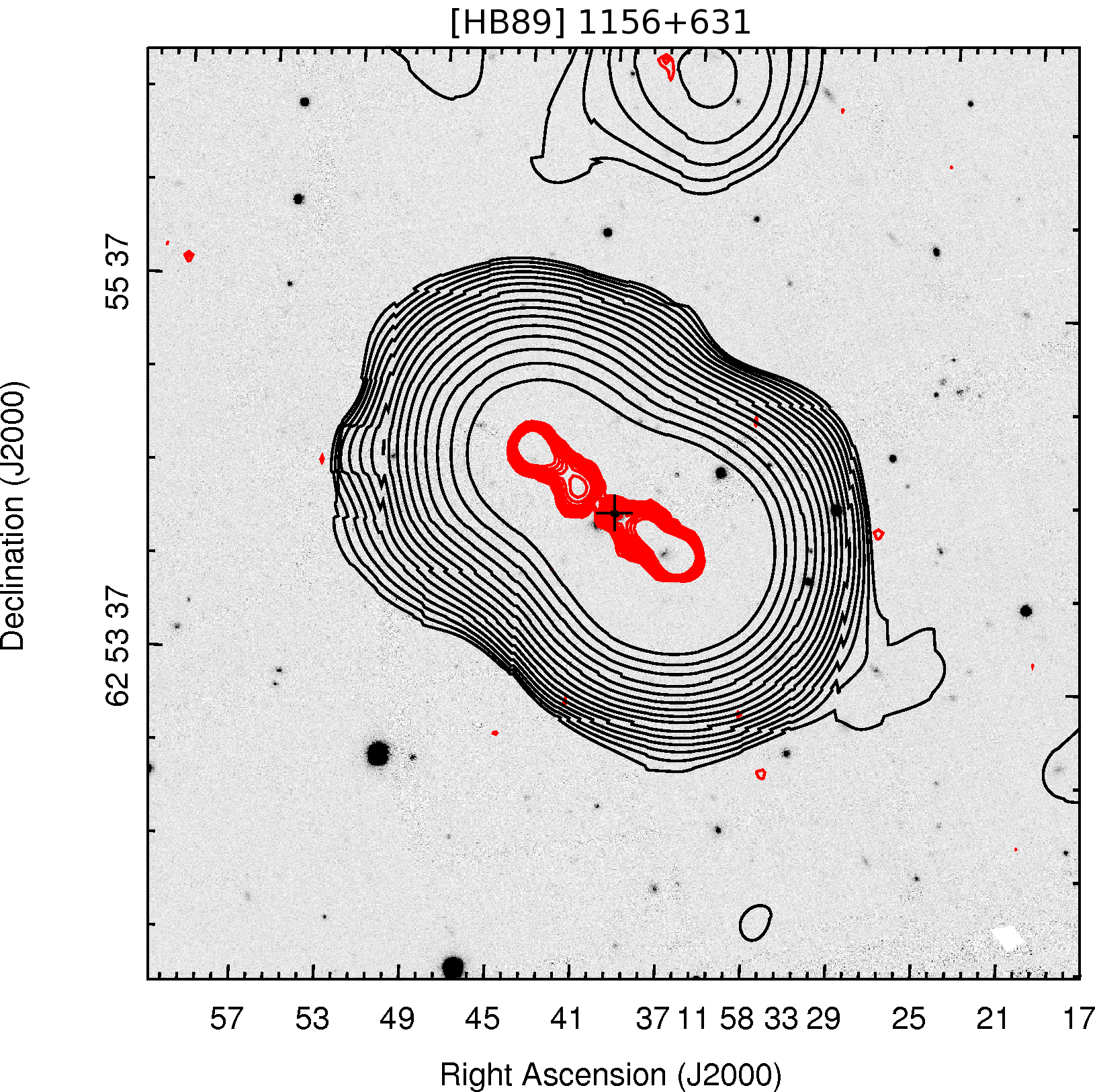}\\
\vspace{0.3cm}
\includegraphics[height=4cm]{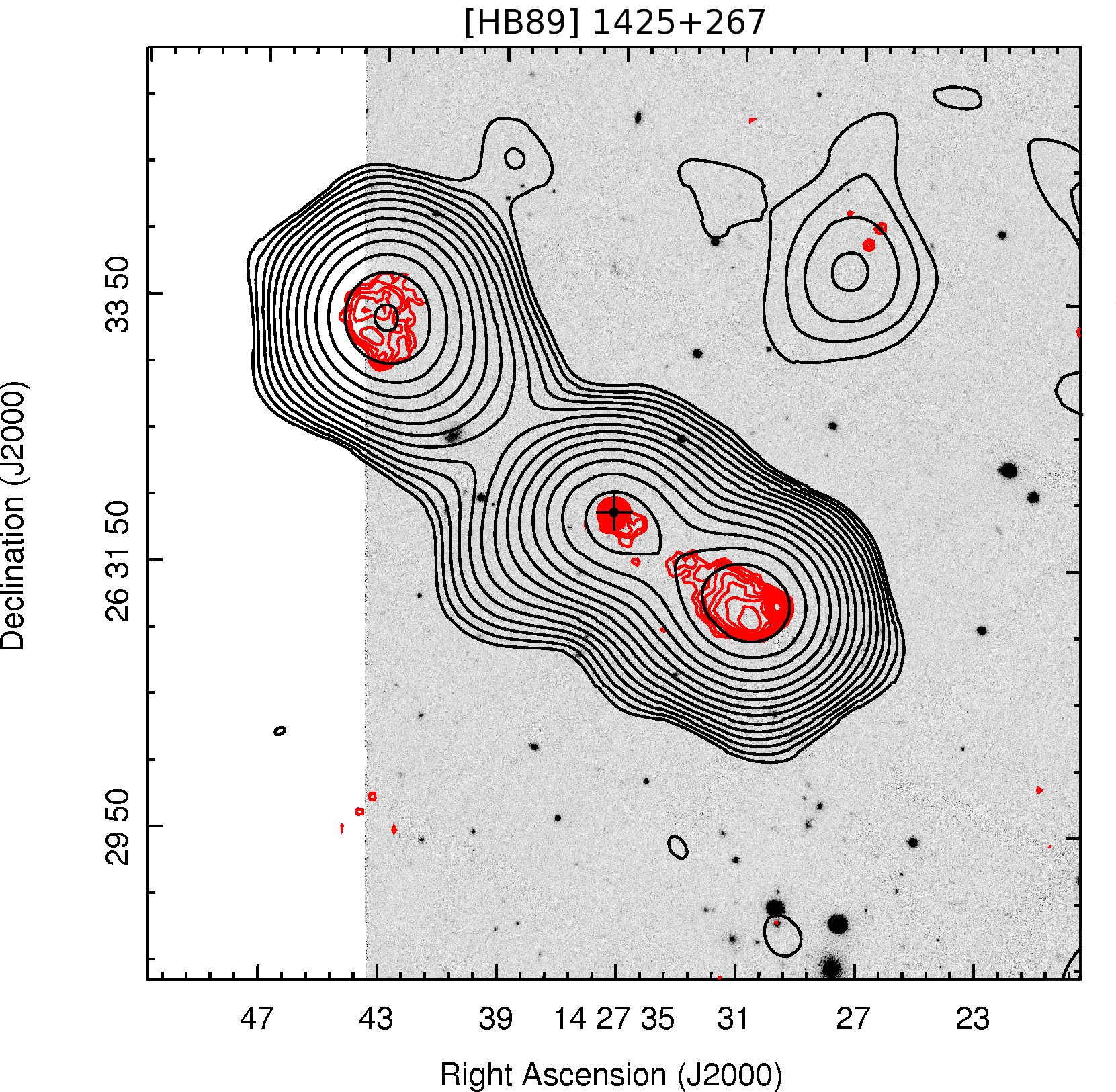}
\includegraphics[height=4cm]{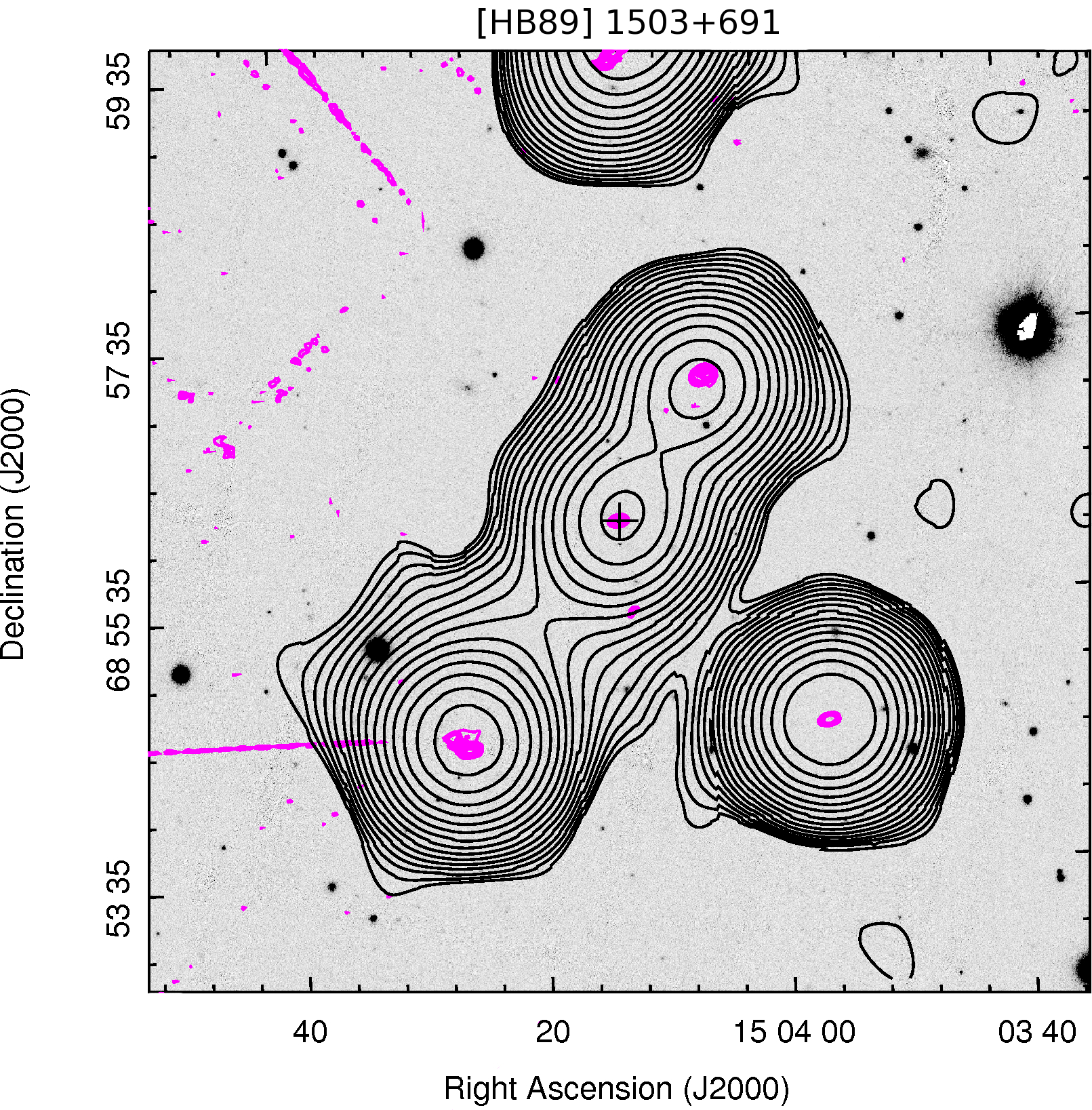}
\includegraphics[height=4cm]{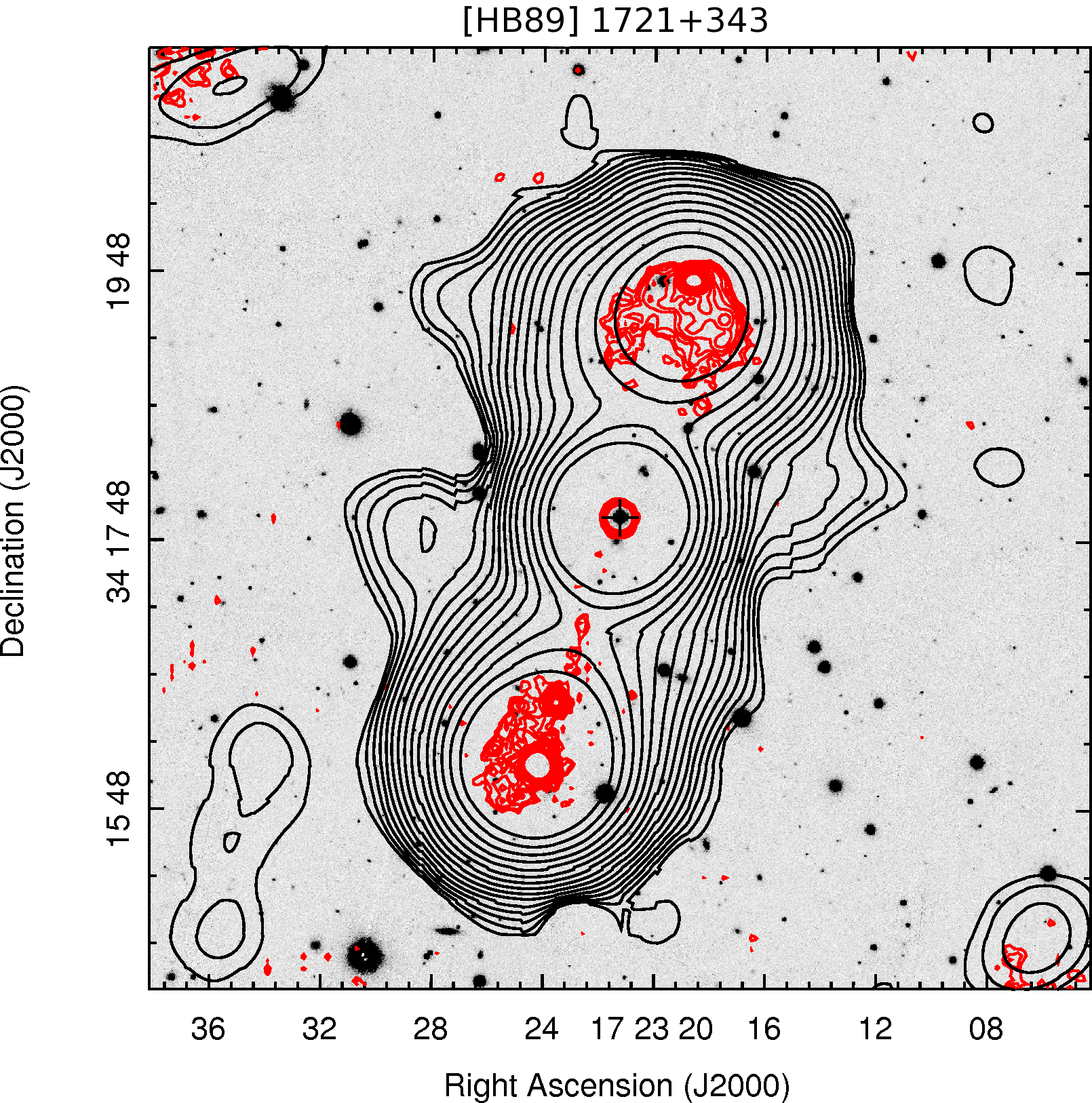}
\includegraphics[height=4cm]{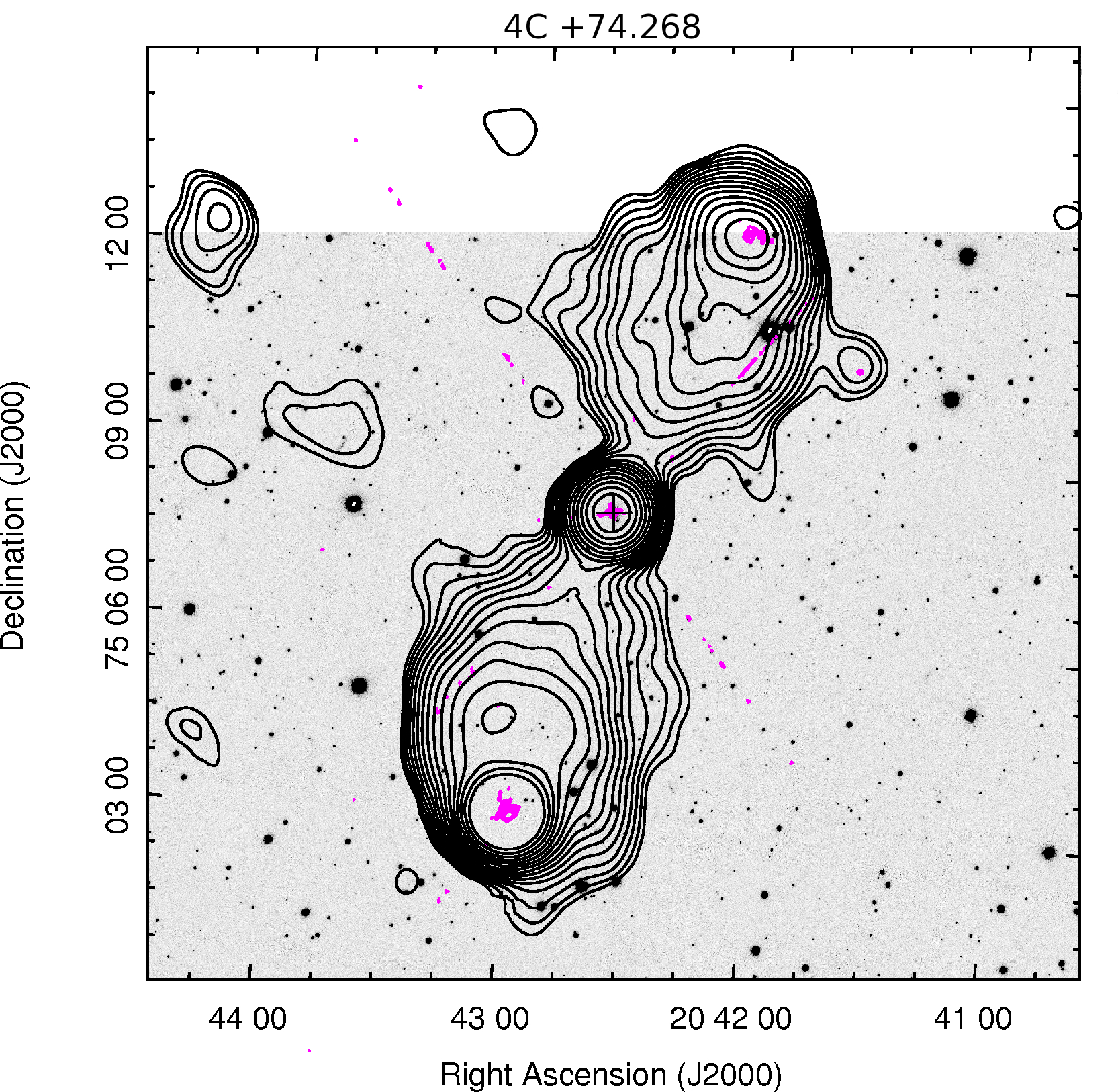}
 
\caption{Radio-optical overlaps of the observed quasars. The 1.4 GHz NRAO Very Large Array Sky Survey (NVSS; \citealt{condon1998}) black contours are overlaid onto the $r$-band Pan-STARRS optical images. The 1.4 GHz Faint Images of the Radio Sky at Twenty-Centimeters (FIRST; \citealt{becker1995}) radio contours are plotted in red, and the 3 GHz Very Large Array Sky Survey (VLASS; \citealt{lacy2020}) contours are in magenta. The contour levels start at nominal 3$\sigma$ rms noise levels of 1.35\,mJy\,beam$^{-1}$, 0.6\,mJy\,beam$^{-1}$, and 0.12\,mJy\,beam$^{-1}$ for the NVSS, FIRST, and VLASS surveys, respectively, and increase by factors of $(\sqrt2)^n$ where $n$ ranges from 0,1,2,3... . The crosses mark the position of the parent QSO. }
\label{fig:radio} 
\end{figure*}

Our sample consists of eight radio-loud quasars with extended radio morphologies of Fanaroff-Riley type II class \citep[][]{fanaroff74}, indicating that the source orientation is close to the sky plane. We selected the sources from the FIRST Bright Quasar Survey \citep{white2000} and the NRAO VLA Sky Survey \citep{condon1998}, based on their mean optical brightness (r$_{\rm mag}<18$) and location on the sky (declination $>$ 10$^\circ$), allowing us to obtain photometry in a few minutes of integration time with small aperture optical telescopes (0.4--0.6 m diameter). Five of the observed quasars are classified as giant radio sources with projected linear sizes exceeding 0.7 Mpc \citep{kuzmicz2012,kuzmicz2021}, whereas the next three are larger than 0.38 Mpc. The basic parameters of the observed QSOs are listed in Table~\ref{tab:sample} and the contour radio maps overlaid on the Pan-STARRS \citep{flewelling2020} optical images are presented in Figure~\ref{fig:radio}.

\begin{deluxetable*}{cccccccccc}
\tabletypesize{\small}
\tablewidth{0pt} 
\tablecaption{Sample properties \label{tab:sample}}
\tablehead{
\colhead{Source Name} & \colhead{RA(J2000)}& \colhead{Dec(J2000)} & \colhead{$z$} & \colhead{$D$} & \colhead{$r_{\rm SDSS}$} & \colhead{$i$}  &\colhead{$\log \rm P_{\rm tot}$}&\colhead{$\log \rm P_{\rm core}$} &Reference \\
\colhead{} & \colhead{} & \colhead{}  & \colhead{} & \colhead{(Mpc)} & \colhead{(mag)} & \colhead{(deg)} &\colhead{W\,Hz$^{-1}$}&\colhead{W\,Hz$^{-1}$} &\colhead{}
} 
\colnumbers
\startdata 
B2 0709$+$37	& 07 13 09.48  & $+$36 56 06.7 & 0.487 & 0.42 & 15.91*& 80 &  25.82& 26.64 & a\\
FBQS J095206.3$+$235245	& 09 52 06.38  & $+$23 52 45.2 & 0.970 & 0.70 & 17.78 & 90  &26.23 & 26.02&  a\\
PG 1004$+$130	& 10 07 26.10  & $+$12 48 56.2 & 0.241 & 0.38 & 15.35  & 67 &  26.26 &24.38  & b \\
$\rm [HB89]$ 1156$+$631	& 11 58 39.90  & $+$62 54 27.9 & 0.592 & 0.39 & 16.33 & 75 &  27.03  & 25.32& b \\
$\rm [HB89]$ 1425$+$267	& 14 27 35.60  & $+$26 32 14.6 & 0.364 & 1.21 & 16.58  & 45 & 26.17 & 25.23 &  a\\
$\rm [HB89]$ 1503$+$691 	& 15 04 12.77  & $+$68 56 12.8 & 0.318 & 0.87 & 17.72* & 81 &  26.13  & 25.52& b \\
$\rm [HB89]$ 1721$+$343 	& 17 23 20.79  & $+$34 17 58.0 & 0.206 & 0.82 & 15.55  & 51 &  26.26  & 25.67& a\\
4C $+$74.268	& 20 42 37.30  & $+$75 08 02.4 & 0.104 & 1.16 & 14.46  & 61 &  25.67   & 24.72& b\\
\enddata
\tablecomments{(1) name, (2--3) J2000.0 source coordinates, (4) spectroscopic redshift, (5) projected linear size of radio structure (assuming H$_0$=71\,km\,s$^{-1}$\,Mpc$^{-1}$, $\Omega_{\rm M}$=0.27, $\Omega_{\rm vac}$=0.73), (6) $r$-band SDSS magnitude, (7) inclination angle from \citet{zola2012} calculated as in \citet{kuzmicz2012}, (8) and (9) total and core radio luminosity at 1.4 GHz from \cite{kuzmicz2012, kuzmicz2021}, (10) reference for source selection: (a) \citet{white2000}; (b) \citet{condon1998}. * the PSF $r$-band magnitudes from the Pan-STARRS data archive. }
\end{deluxetable*}

\section{Observations and data reduction}

The optical monitoring of the source sample began in March 2009 as part of the Krak{\'o}w Quasar Monitoring program.
 Observations have been gathered mainly from two telescopes: the 60\,cm one at Mt. Suhora Observatory of the Pedagogical University (65.7\%) and the 50\,cm Cassegrain telescope of the Astronomical Observatory of the Jagiellonian University (21.3\%). To avoid long gaps due to long periods of bad weather at the two sites, supplementary observations were carried out using robotic telescopes operated by the SKYNET Robotic Telescope Network (\citealt{zola2021}) that constituted 13\% of the observations. In particular, we used the 50\,cm CDK telescope of the Astronomical Observatory of the Jagiellonian University, the 40\,cm telescope of the Dark Sky Observatory (DSO), the 60\,cm Rapid Response Robotic Telescope (RRRT) of the Fan Mountain Observatory, the 60 cm and 100 cm telescope of the Yerkes Observatory (YERKES), the 40\,cm telescope of the Northern Skies Observatory (NSO), 40\,cm telescope of the Montana Learning Center (MLC), 40\,cm telescope of the Perth Observatory (RCOP), 40\,cm telescope of the Mars Desert Research Station (MDRS), and the 40\,cm telescope of the Dolomiti Astronomical Observatory (DAO). All telescopes are equipped with CCD cameras and wide-band filters. The quasars in our sample were observed in the $R$-band taking $\sim$10 frames per source each clear night. The integration times varied between 1 and 3 minutes depending on the quasar brightness and weather conditions. The scientific frames taken with the Mt. Suhora and Krakow Cassegrain telescopes were corrected for bias, dark, and flat field in the usual manner using the IRAF package\footnote{\url{http://iraf-community.github.io}}. Series of bias and dark images were taken before or after each night, while flat field images were mostly taken on the twilight sky. Calibration of images taken by the Skynet telescopes was performed by the network software pipepline. For each calibrated quasar frame, we performed differential aperture photometry using the {\it CMunipack}\footnote{\url{http://c-munipack.sourceforge.net}} program by selecting two stars (a comparison and a check star) falling on the same CCD frame. Out of this, the comparison star was used to derive the differential magnitude of the quasar while the check star was used to derive the differential magnitude of the comparison star. Lack of variability in the differential light curves of the comparison star ensured that the variations seen in the quasar differential light curves are intrinsic and not due to the variability of the comparison star itself.\\
 
The resultant differential magnitudes were averaged over the night, thus we obtained one mean measurement per night. The gathered light curves are presented in Figure~\ref{lc1}. In our analysis, we used comparison and check stars listed in the Appendix~\ref{appendix1}. The light curves for each quasar are available online as supplemental material (for details, see the Appendix~\ref{appendix2}).
Each target was observed 2-4 times per month, occasionally with a denser coverage, uniformly covering a $\sim$13-year period. Further uniformity was secured by using the same comparison star in frames taken at each site, and the reduction of all data was carried out by a single person (AK). 
We note that we could not monitor quasar PG\,1004+130 for about four months each year because it had an altitude too low to allow monitoring from the (northern) Polish observing sites.

\begin{figure*}[htbp!] 
\centering 
\includegraphics[height=5.3cm]{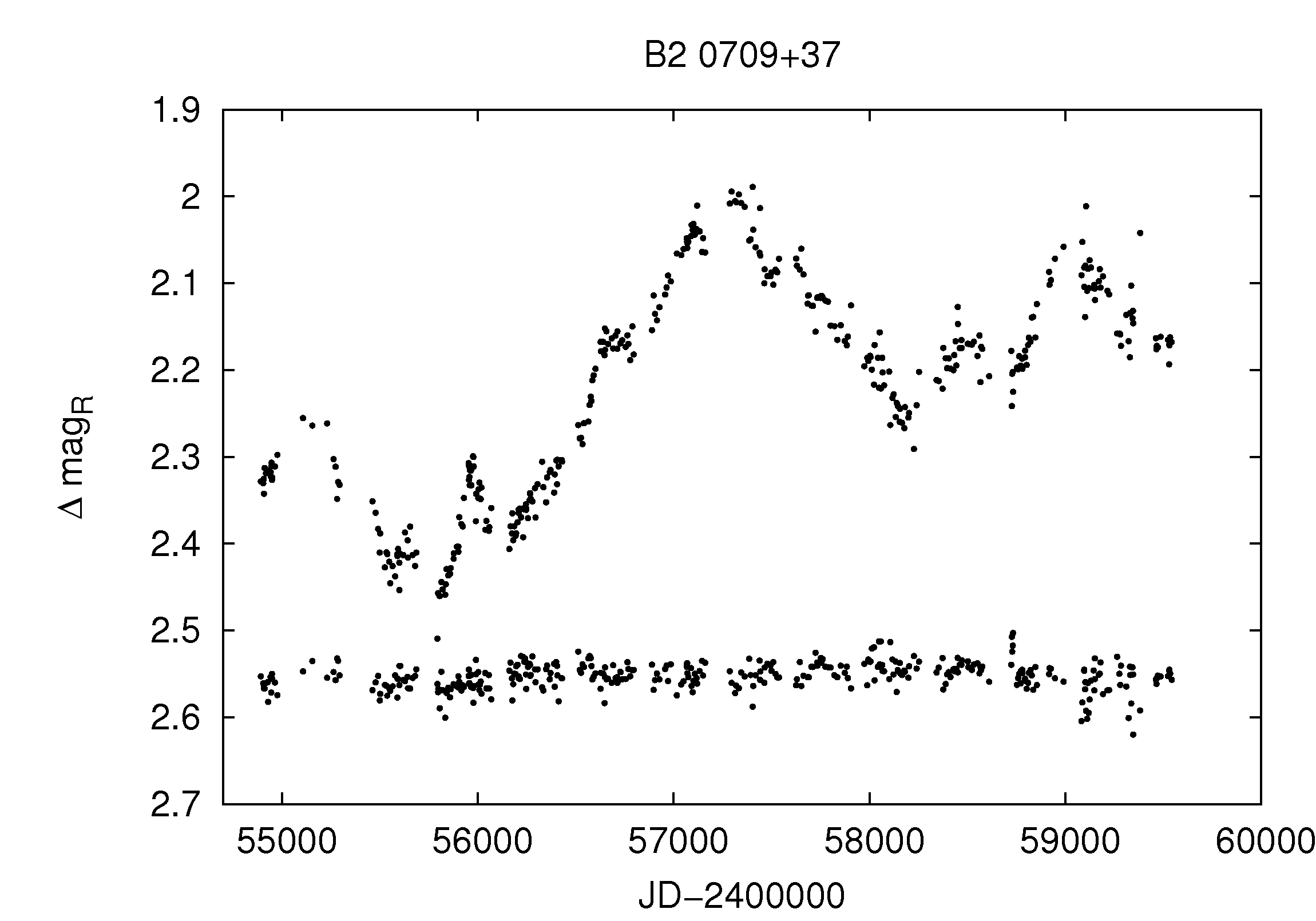}
\includegraphics[height=5.3cm]{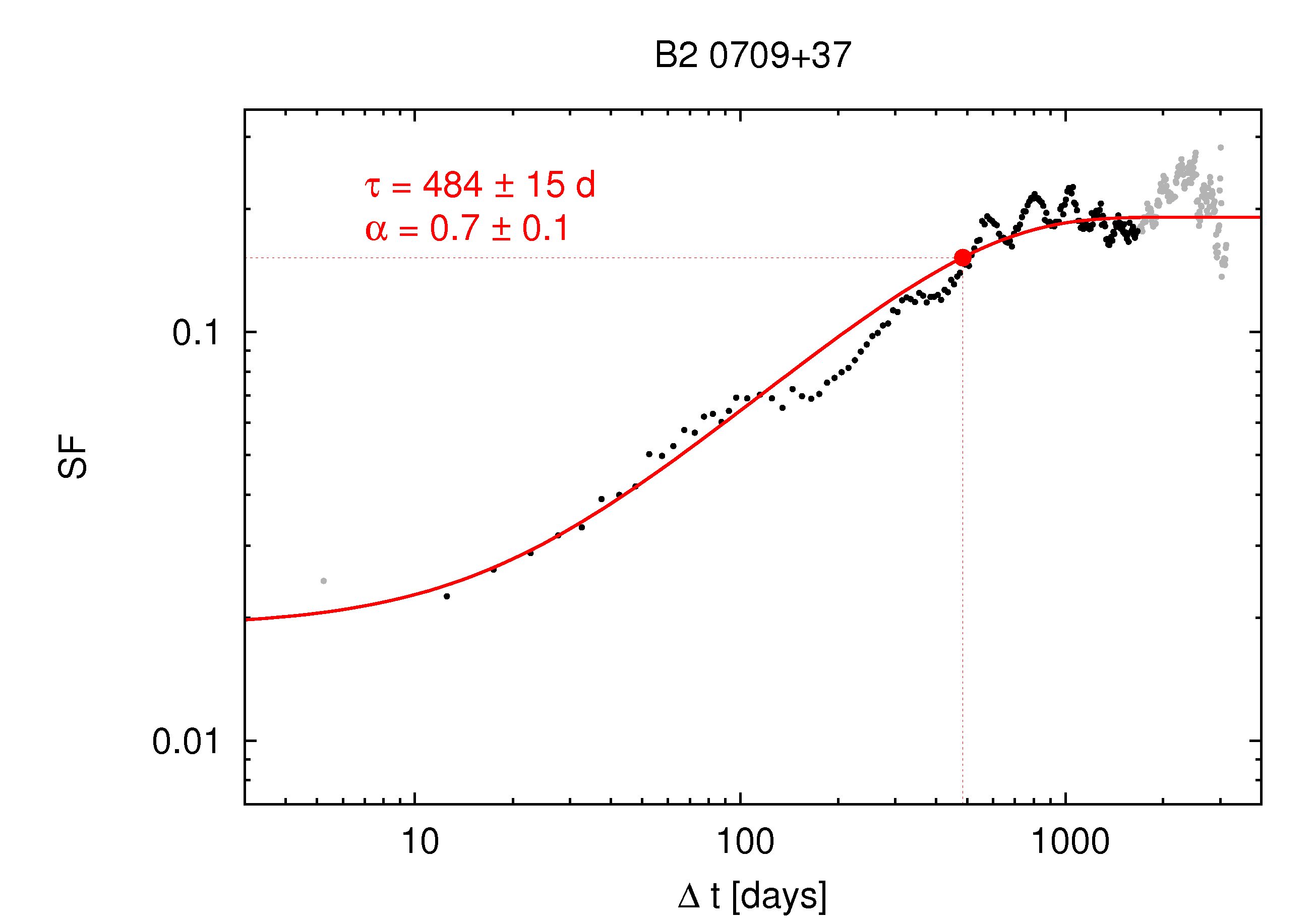}\\
\vspace{0.2cm}
\includegraphics[height=5.3cm]{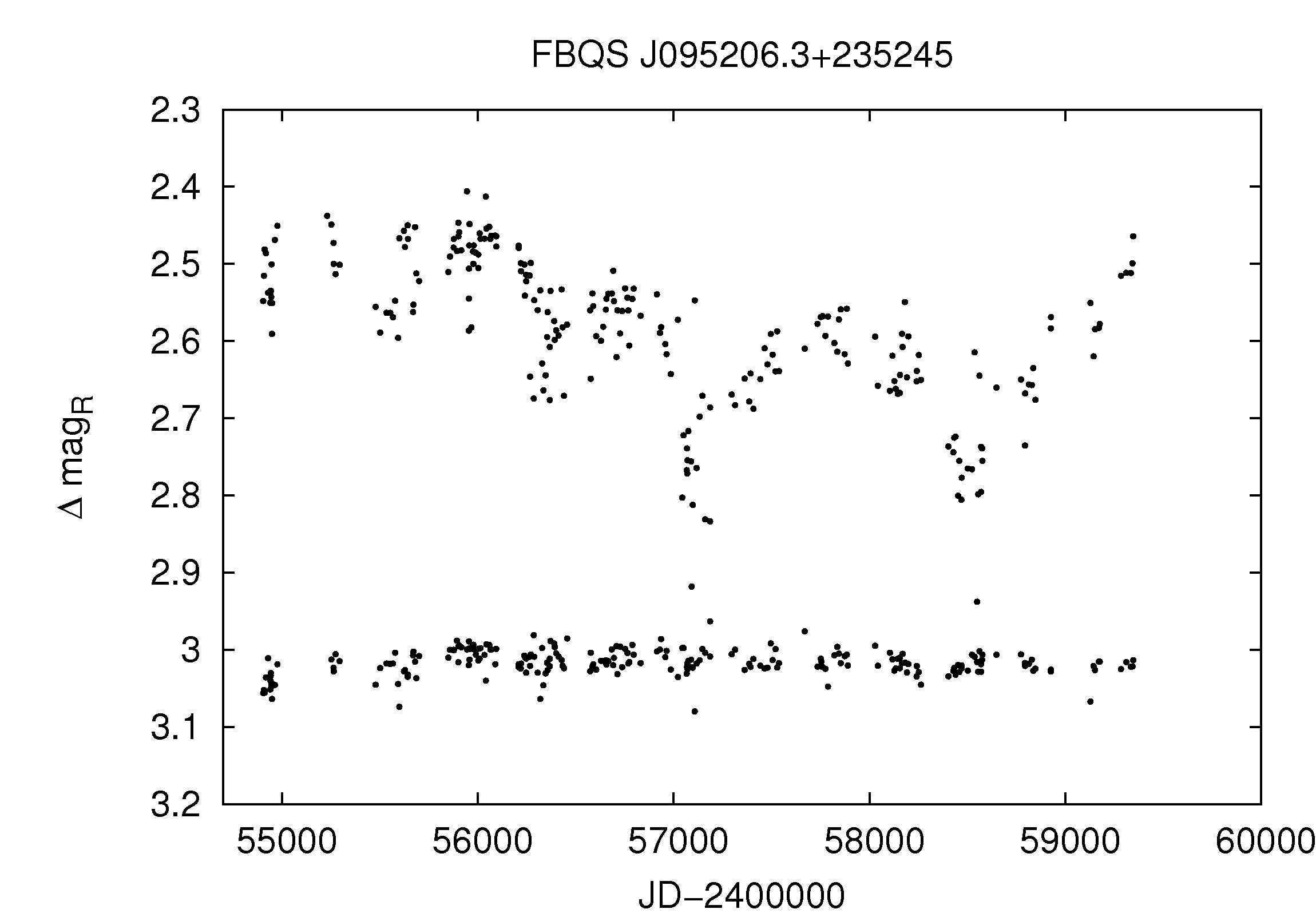}
\includegraphics[height=5.3cm]{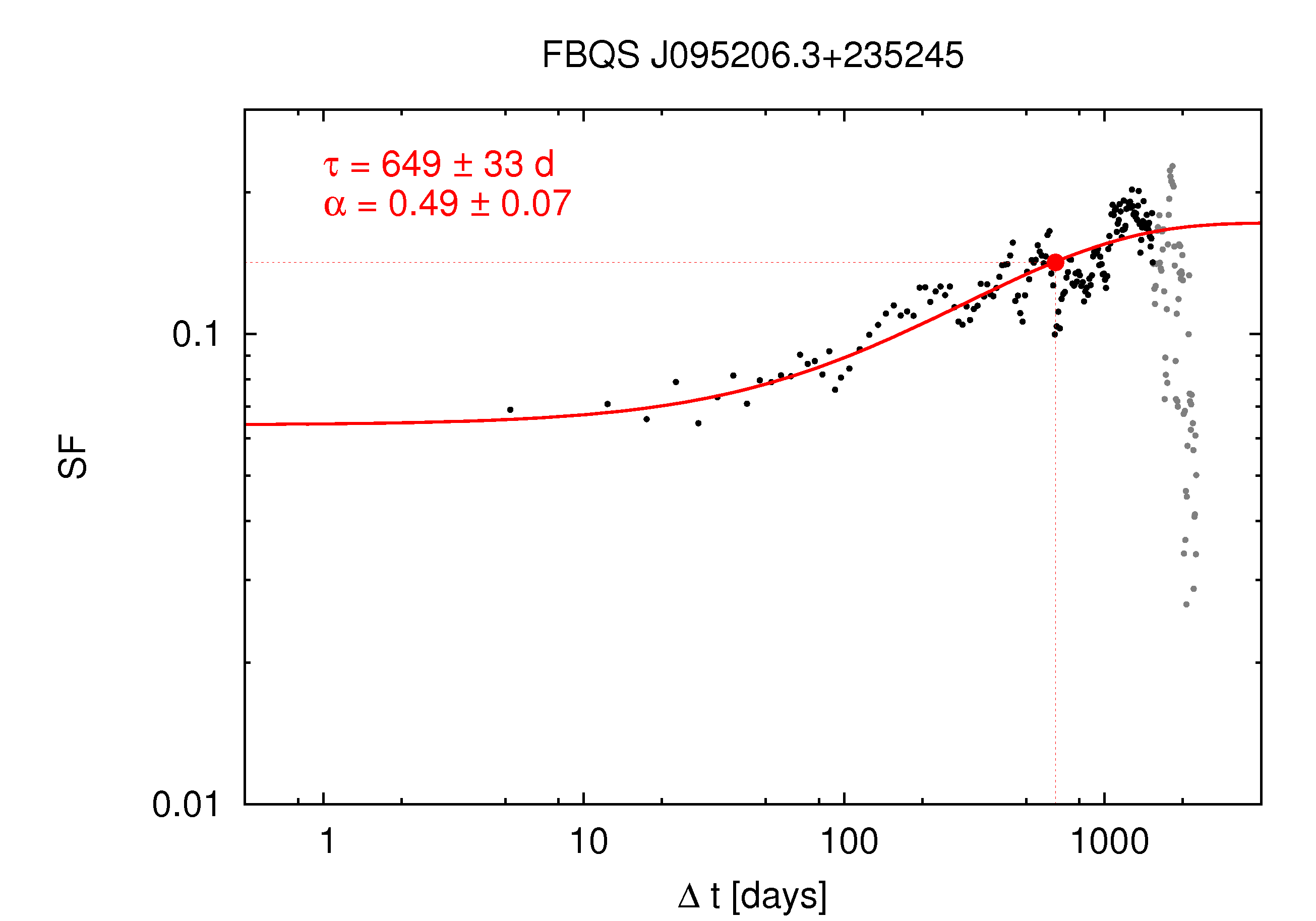}\\
\vspace{0.2cm}
\includegraphics[height=5.3cm]{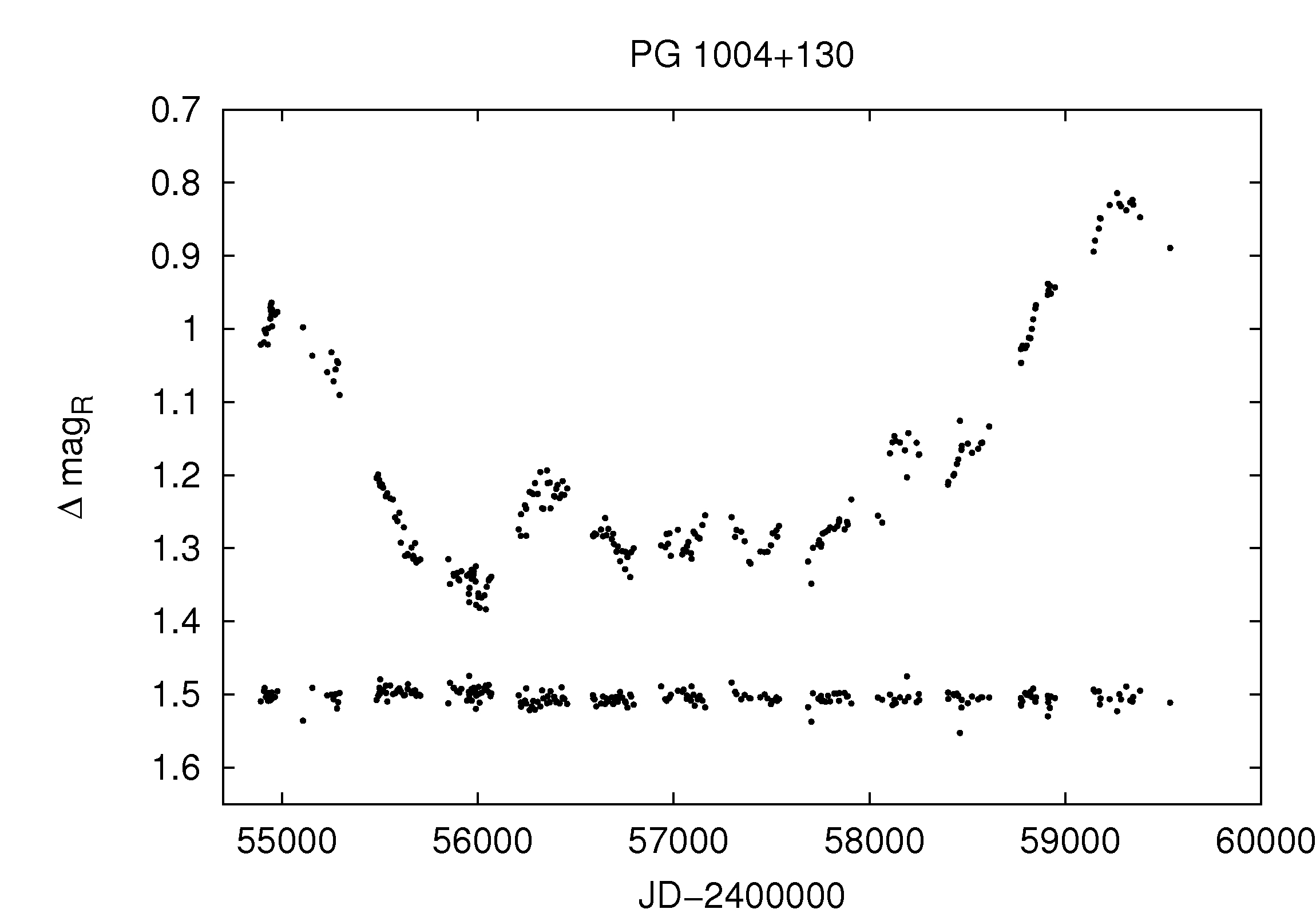}
\includegraphics[height=5.3cm]{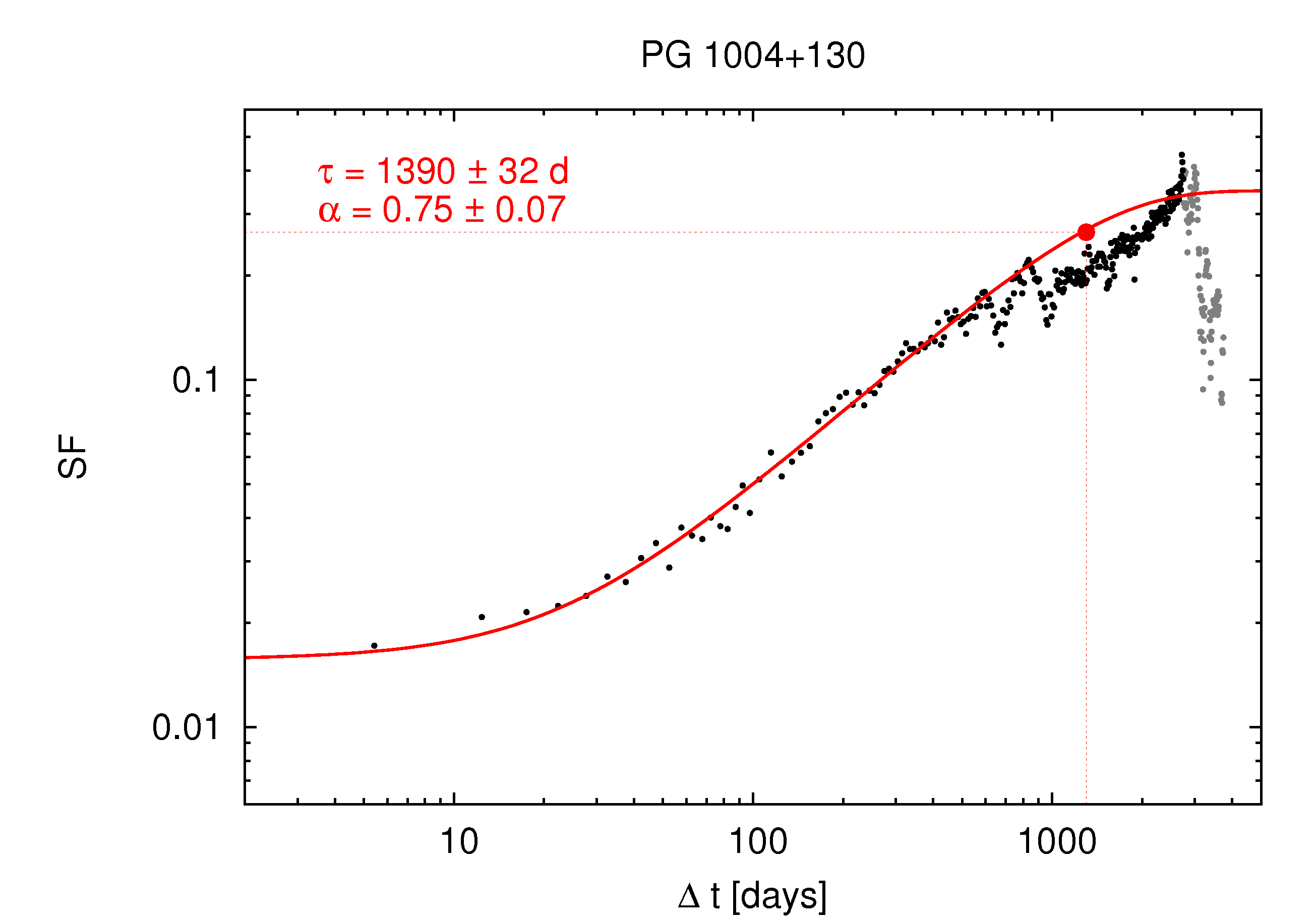}\\
\vspace{0.2cm}
\includegraphics[height=5.3cm]{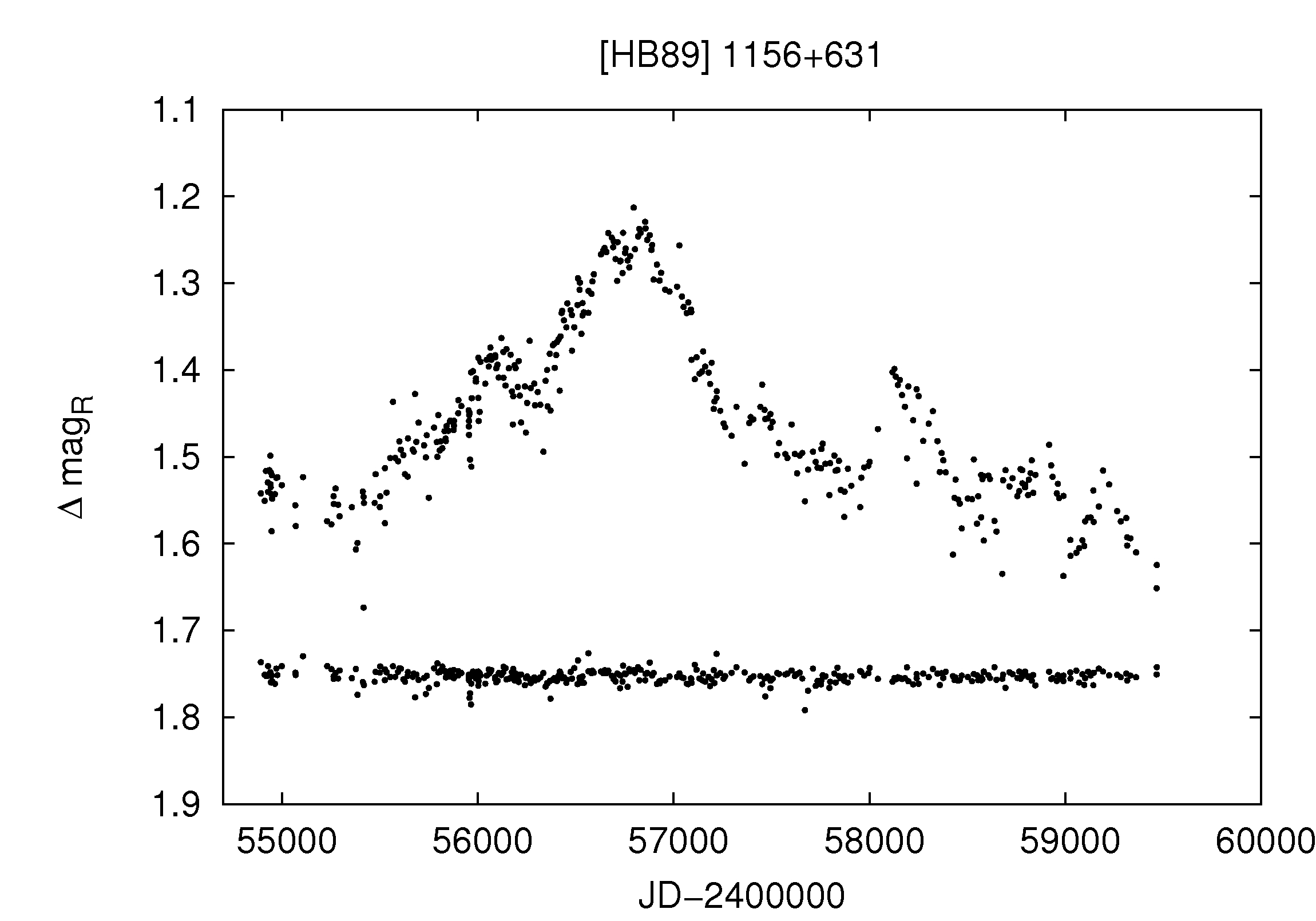}
\includegraphics[height=5.3cm]{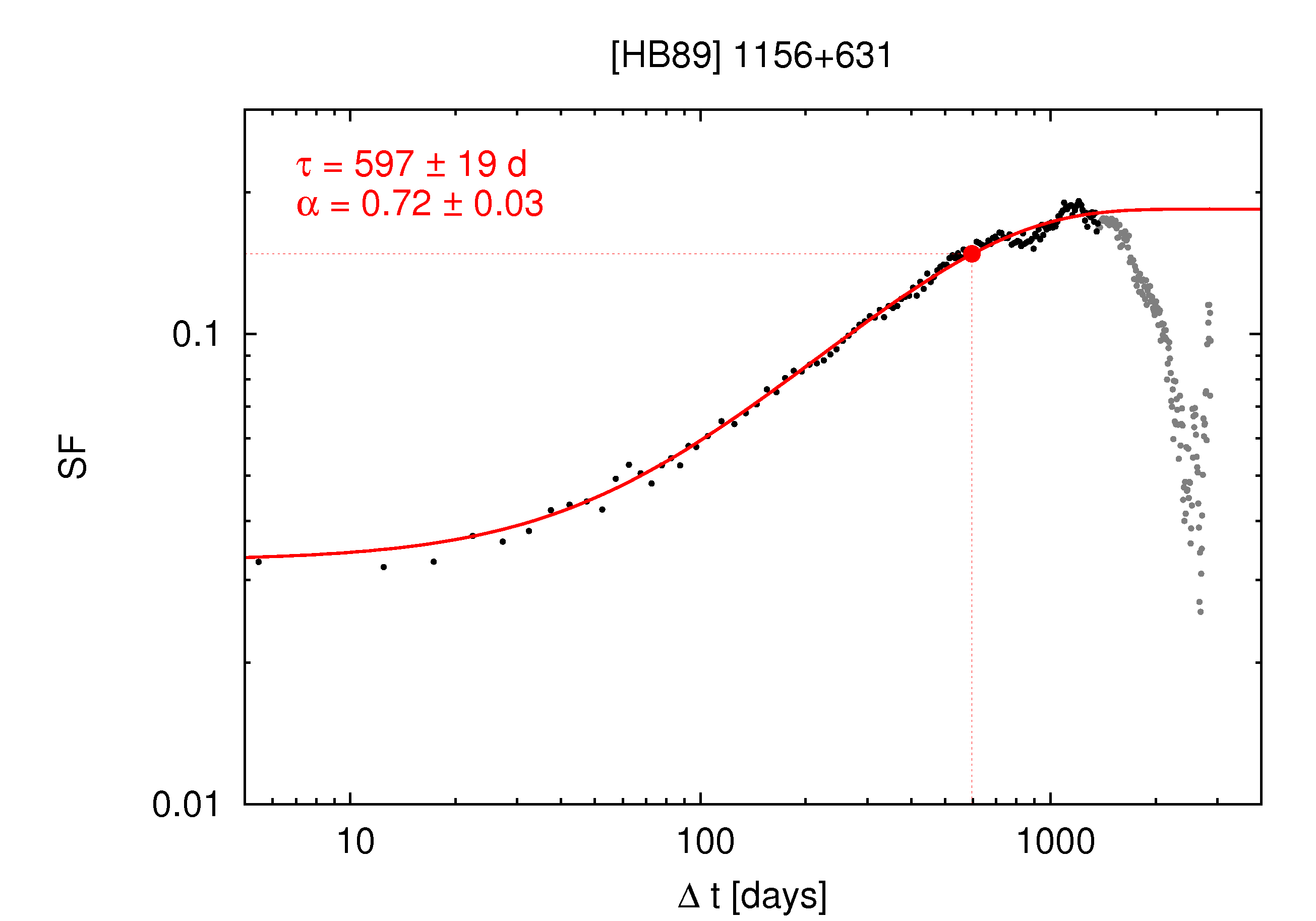}\\
\caption{{\bf Left panels:} $R$-band light curves for B2 0709$+$37, FBQS J095206.3$+$235245, PG 1004$+$130 and [HB89] 1156$+$631 quasars. The quasar variability is expressed as the relative magnitude changes with respect to the comparison star. In the bottom part of each graph, we plot the relative magnitudes between the comparison and the check star. {\bf Right panels:} Corresponding structure functions (Section~\ref{sec:sf}). For large time lags $\Delta$t (grey points), the sampling of SF is poor due to the finite range of time series, and the characteristic plateau is not visible. Therefore, we do not take them into account when fitting the four-parameter SF function plotted in red. The red point corresponds to the decorrelation time scale $\tau$ (Eq.~\ref{eq:sfnoise}) . } 
\label{lc1} 
\end{figure*}

\addtocounter{figure}{-1}
\begin{figure*}[htbp!] 
\centering 
\vspace{0.2cm}
\includegraphics[height=5.3cm]{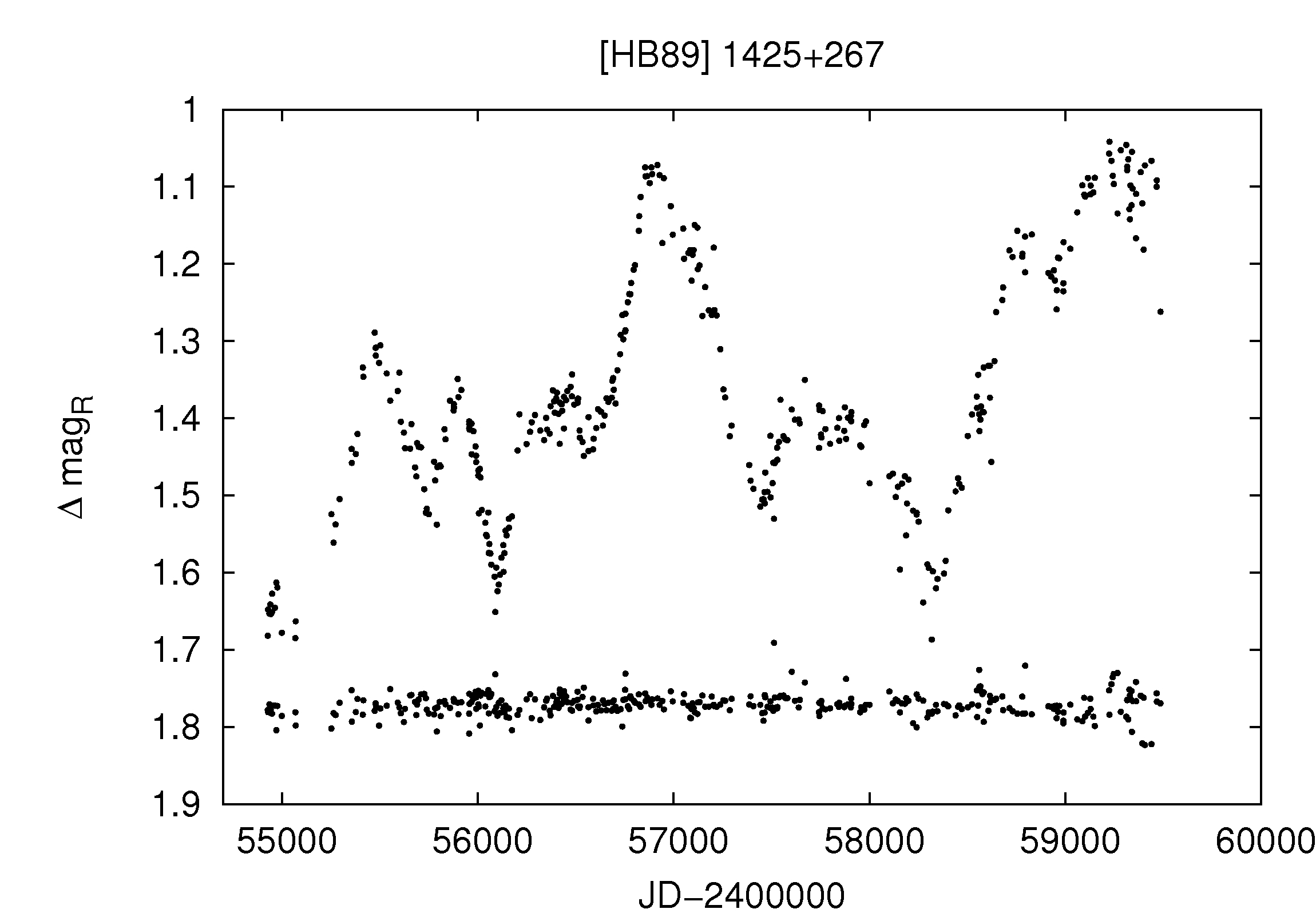}
\includegraphics[height=5.3cm]{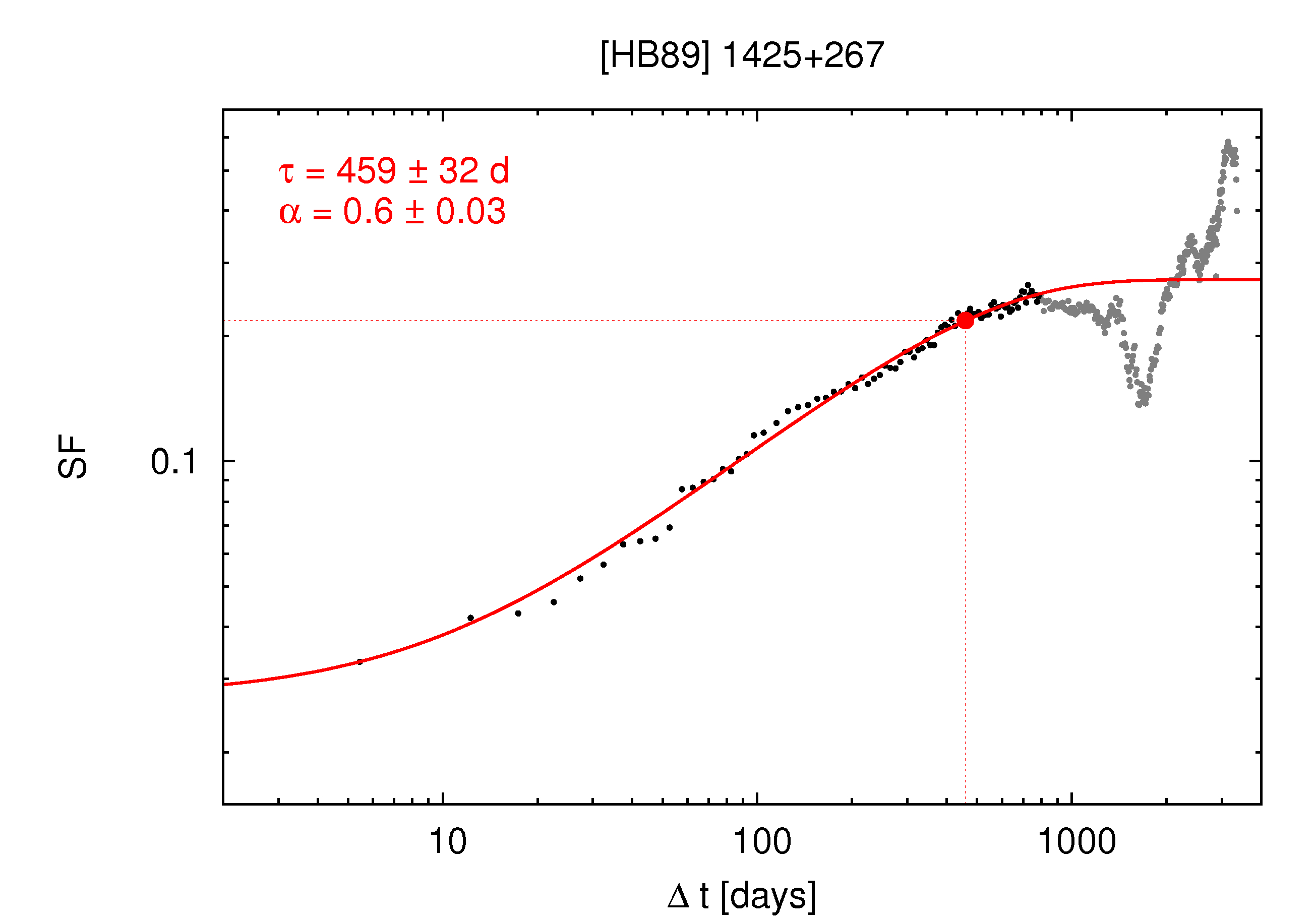}\\
\vspace{0.2cm}
\includegraphics[height=5.3cm]{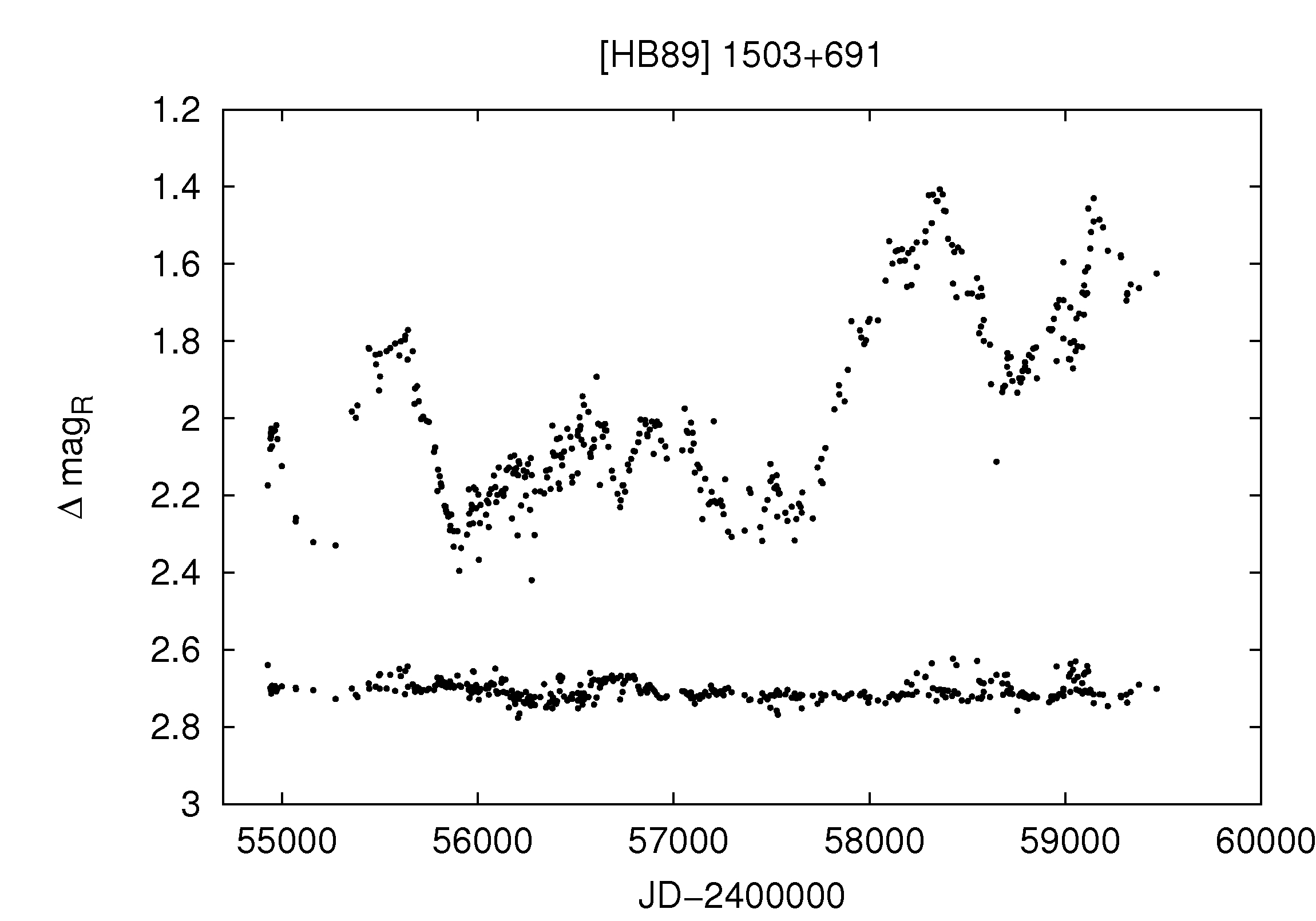}
\includegraphics[height=5.3cm]{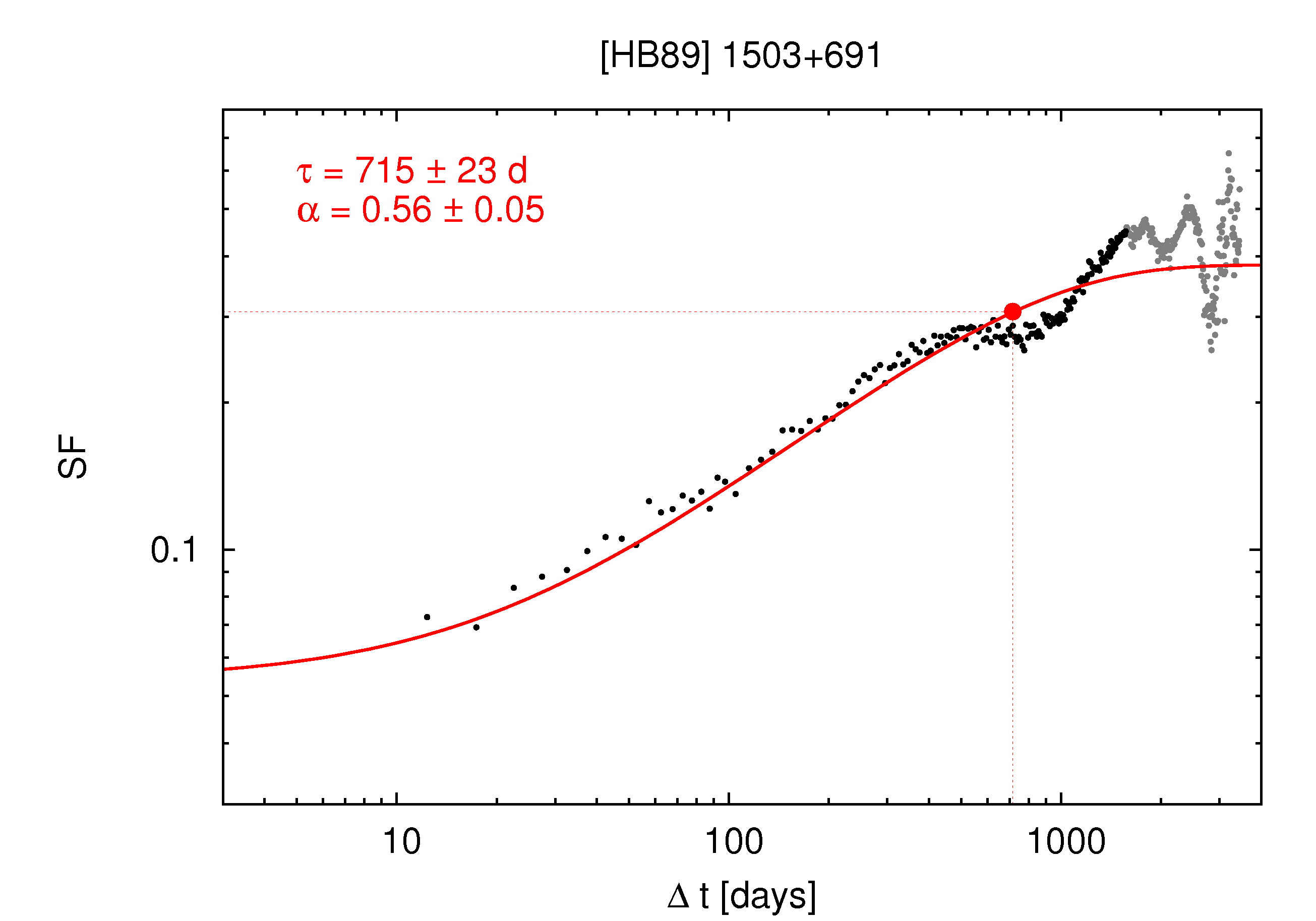}\\
\vspace{0.2cm}
\includegraphics[height=5.3cm]{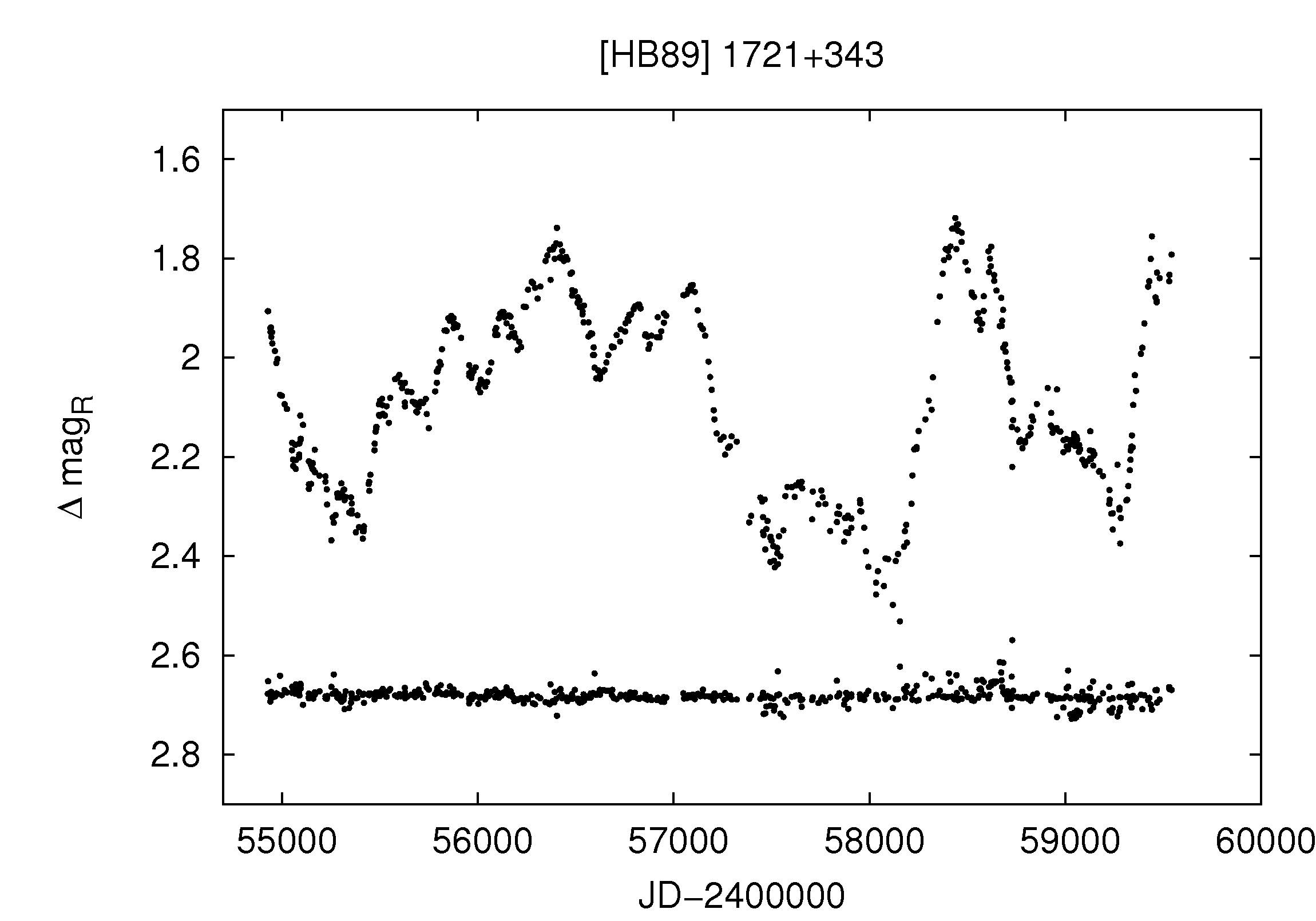}
\includegraphics[height=5.3cm]{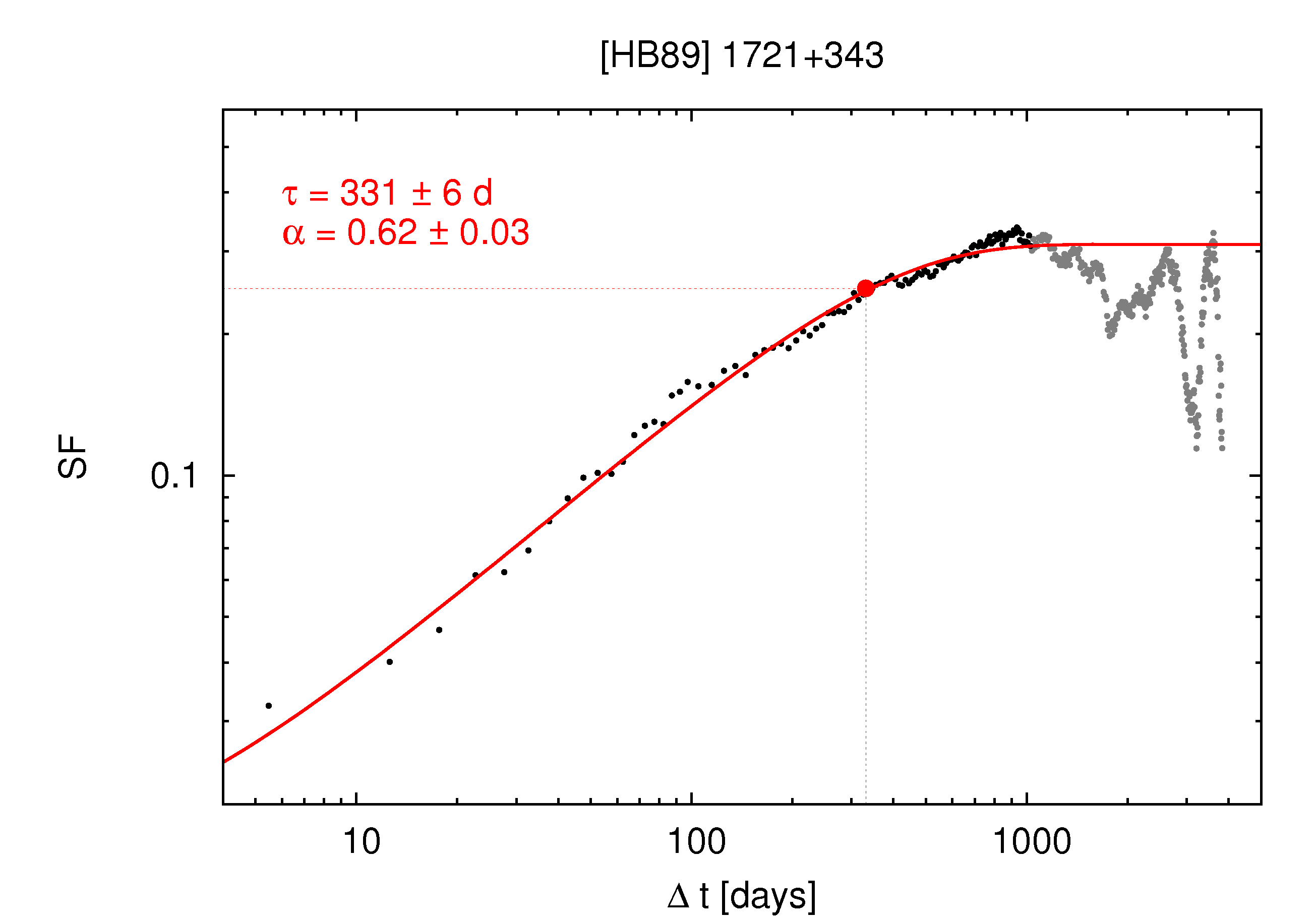}\\
\vspace{0.2cm}
\includegraphics[height=5.3cm]{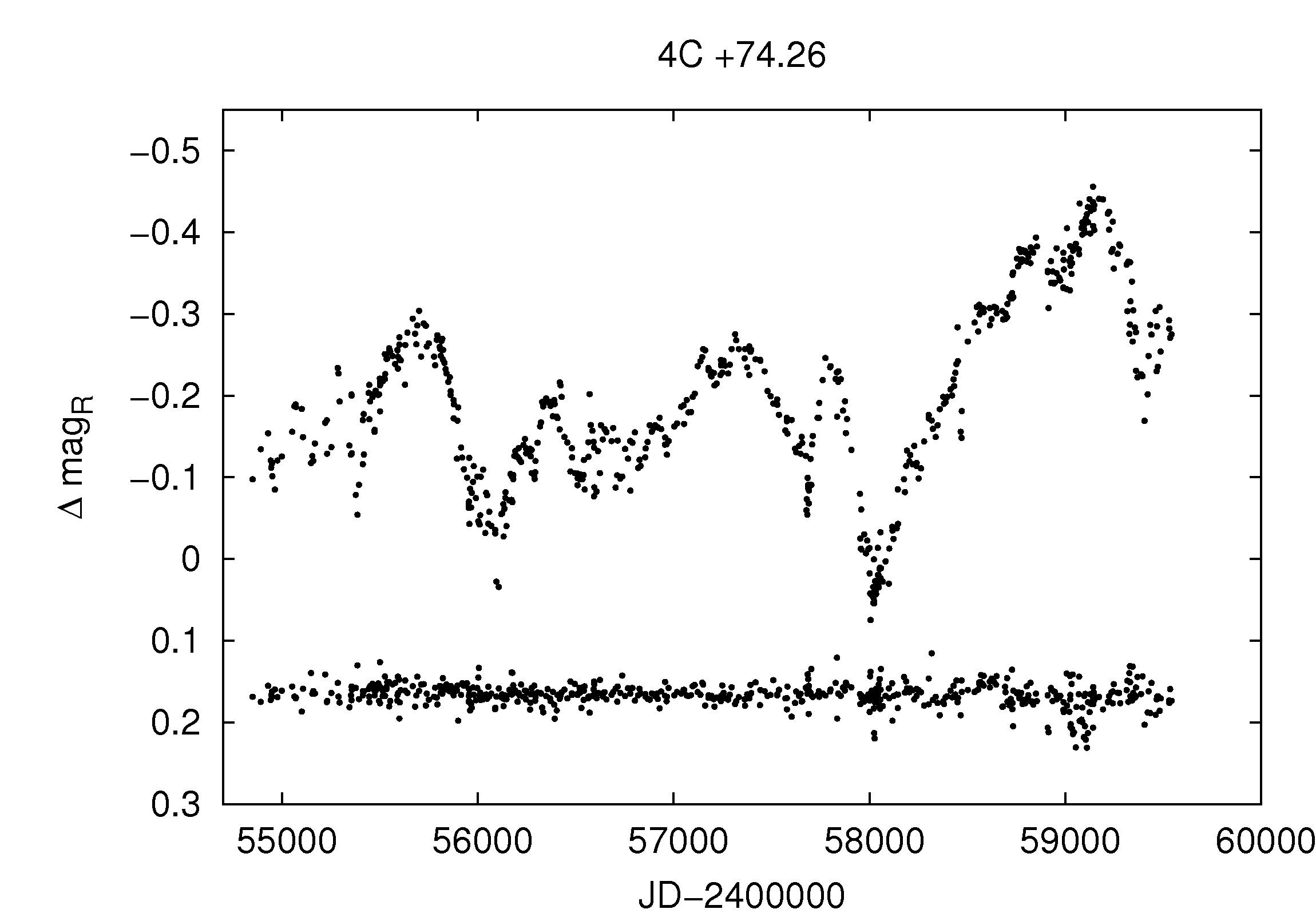}
\includegraphics[height=5.3cm]{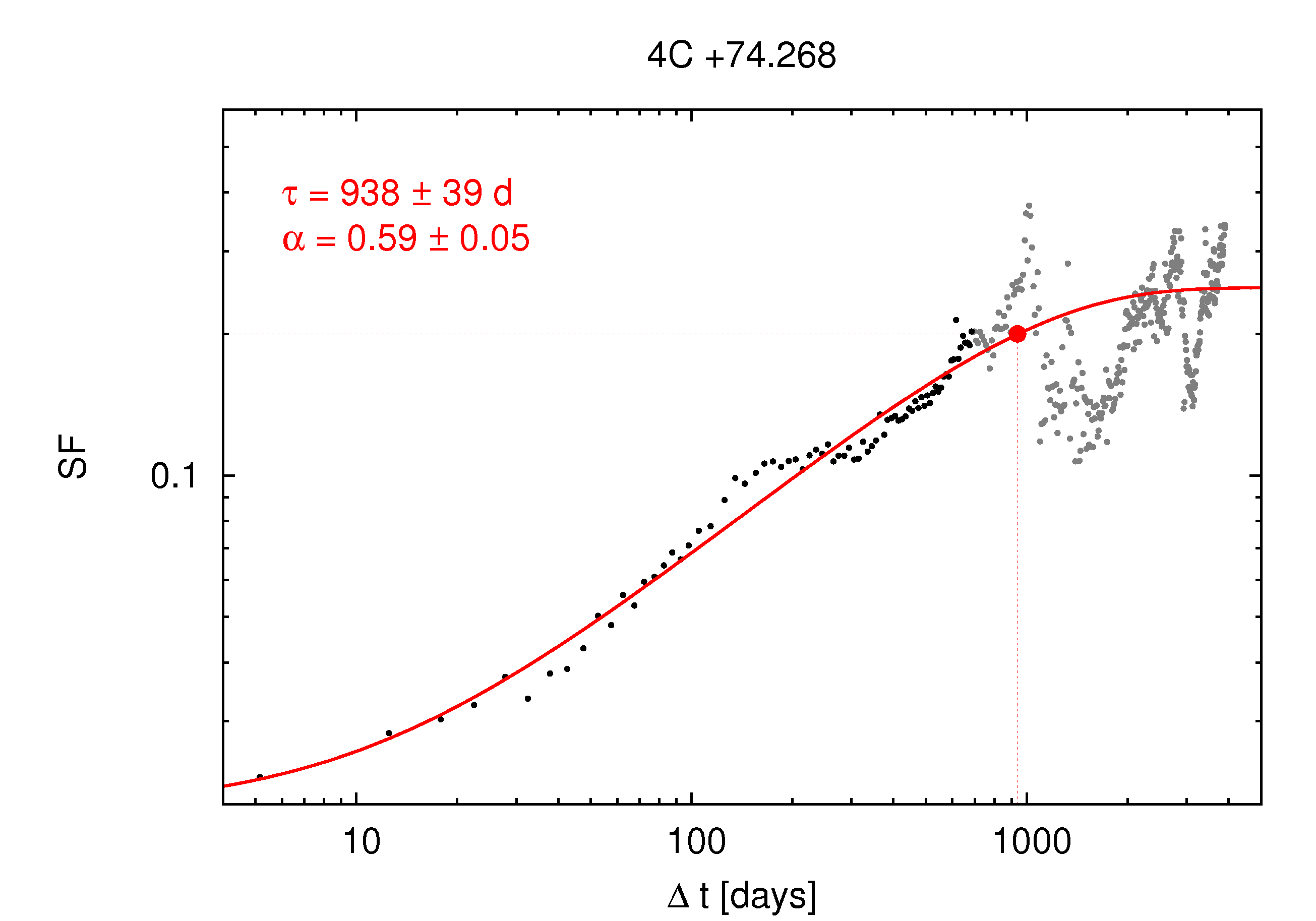}\\
\caption{\textit{continued:} $R$-band light curves for [HB89] 1425$+$267, [HB89] 1503$+$691 , [HB89] 1721$+$343 and 4C $+$74.268 quasars with corresponding structure functions. } 
\end{figure*}

\section{Analysis methods}

We use two alternative methods to characterize the variability of our sources: structure function (SF) analysis, which falls into the category of the time-domain methods, and power spectral density (PSD) analysis, which falls into the category of Fourier-domain methods. These alternative approaches are needed to understand inherent biases that can arise when interpreting the results of variability analysis methods due to the nature of the time series itself. The ground-based monitoring datasets at optical wavelengths are usually irregularly sampled, as they contain periodic gaps due to daytime and observing seasons, as well as aperiodic gaps due to observing constraints on a given night, leading to unevenly sampled time series. It has been shown that the application of the SF and PSD methods in unevenly sampled time series is known to produce false features in their results, and therefore such features should be treated with caution \citep[see, in this regard; ][]{emmanoulopoulos2010, goyal2020, goyal2022}. In the following sections, we discuss the strengths and weaknesses of these methods.

\subsection{Structure function}\label{sec:sf}

The SF provides quantitative measures of how strongly the flux changes as a function of the time interval, i.e., it measures the mean value of the magnitude difference for measurements separated by a given time interval. Several approaches to structure function calculations are employed in the literature (e.g., \citealt{emmanoulopoulos2010, macLeod2011, kozlowski2016}); however, in this work we use the formalism described by \citet{simonetti1985} with SF defined as: 
\begin{equation}\label{eq:sf}
\rm SF(\Delta t)^2=\frac{1}{N}\sum_{i,j>i}^N{(m_i-m_j)^2}
\end{equation}\\
where the sum is over all $N$ magnitude measurements that are separated by some time lag $\Delta t$. To correct for the observed frame variability to the rest frame, we use $\Delta t=|t_i-t_j|(1+z)^{-1}$ in Eq.~\ref{eq:sf}.  
SF is commonly characterized in terms of its slope $\alpha$, where SF$^2$($\Delta\,t$)\,$\propto$\, $\Delta\,t^{\alpha}$. Interpreting the specific variability characteristics based on the shape of the SF and its properties becomes sometimes complicated due to the complex shape of the SF. However, some properties can provide information about typical timescales of observed variability and can be used to search for periodicities. A characteristic timescale of variability, defined as the time interval between the maximum and the neighboring minimum in a light curve, manifests itself as the maximum of the SF, while the periodicity in the light curve is visible as a minimum of SF \citep[e.g., ][]{heidt1996}.  Furthermore, the slope of the SF characterizes the process underlying the variability. Based on theoretical models of AGN variability, \citet{hawkins2002} determined the predicted slopes of the structure function, which are equal to 0.83, 0.44, and 0.25 for the starburst, accretion disc instability and microlensing models, respectively. \\
Eq.~\ref{eq:sf} does not include the contribution to the SF values due to measurement noise, which should be subtracted to obtain the ``true SF'' \citep[for a comprehensive review, see][] {kozlowski2016}. Therefore, to obtain the correct results that characterize the variability, we fit the four-parameter function to the SF calculated by Eq.~\ref{eq:sfnoise} (\citealt{kozlowski2016}):
\begin{equation}\label{eq:sfnoise}
\rm SF^2(\Delta t)=\rm SF^2_{\infty}[1-\rm exp(-|\Delta t|/\tau)^{2\alpha})]+2\sigma^2
\end{equation}\\        
where SF$_{\infty}$ is the variance on long timescales, $\tau$ is the decorrelation timescale at which the SF changes slope -- the characteristic timescale of variability, $\alpha$ is the slope of the SF, and $\sigma$ is the noise term. The obtained SFs with the fitted functions calculated for our quasar light curves are presented in Figure~\ref{lc1}, and the basic information on the light curves and the SF fitted parameters is listed in Table~\ref{tab:sf}. 

\begin{deluxetable*}{cccccc}
\tabletypesize{\small}
\tablewidth{0pt} 
\tablecaption{Light curve parameters and the results of structure function analysis \label{tab:sf}}
\tablehead{
\colhead{Source Name}  & \colhead{N$_{\rm lcp}$} &  rms &   \colhead{$\alpha$} & \colhead{$\tau$} & \colhead{$\Delta m$} \\
\colhead{} & \colhead{} & \colhead{(mag)} & \colhead{}  & \colhead{(day)} & \colhead{(mag)} 
} 
\colnumbers
\startdata 
B2 0709$+$37	        & 378 & 0.028   & 0.70 $\pm$ 0.10  & 484 $\pm$ 15  & 0.46$\pm$0.02  \\
FBQS J095206.3$+$235245	& 223 & 0.090   & 0.49 $\pm$ 0.07  & 649 $\pm$ 33  & 0.51$\pm$0.08   \\
PG 1004$+$130       	& 249 & 0.018    & 0.75 $\pm$ 0.07  & 1390$\pm$ 32  & 0.56$\pm$0.02   \\
$\rm [HB89]$ 1156$+$631	& 365 & 0.031   & 0.72 $\pm$ 0.03  & 597 $\pm$ 19  & 0.44$\pm$0.03   \\
$\rm [HB89]$ 1425$+$267	& 353 & 0.040   & 0.60 $\pm$ 0.03  & 459 $\pm$ 32  & 0.64$\pm$0.03  \\
$\rm [HB89]$ 1503$+$691 & 369 & 0.069   & 0.56 $\pm$ 0.05  & 715 $\pm$ 23  & 0.97$\pm$0.06  \\
$\rm [HB89]$ 1721$+$343 & 505 & 0.024   & 0.62 $\pm$ 0.03  & 331 $\pm$ 6  & 0.81$\pm$0.02  \\
4C $+$74.26	            & 562 & 0.012   & 0.59 $\pm$ 0.05  & 938 $\pm$ 39  & 0.53$\pm$0.01 
\enddata
\tablecomments{(1) name, (2) number of data points in the light curve, (3)  mean error of differential quasar magnitude, (4) slope of the structure function, (5) characteristic variability timescale, (6) observed variability amplitude.
 }
\end{deluxetable*}

The SF slopes obtained for the studied quasars are relatively steep. According to the model predictions (\citealt{hawkins2002}) almost all have values between 0.44 $\pm$ 0.03 and 0.83 $\pm$ 0.08  corresponding to the theoretical values of the SF slopes for the accretion disc instability model and the starburst model, respectively. Therefore, it is difficult to unambiguously designate the mechanism of variability in the case of our quasar sample. However, the starburst model is sufficient to explain the magnitude changes in low-luminosity AGNs. It becomes problematic to explain magnitude changes in moderate and most luminous quasars (e.g. \citealt{hawkins2002}). It should also be stressed that, as all of the observed quasars have radio lobes lying close to the sky plane, the optical variability mechanisms are expected to originate at the accretion disc. Nevertheless, it cannot be excluded that observed variability is in some extent modified by the Doppler-boosted emission from radio jets. The large-scale radio emission manifested by the radio lobes (Figure \ref{fig:radio}) was generated in a different epoch from the optical variability observed during the last 13 years. The orientation of the quasar could change during that time, and now the radio jets can propagate closer to the line of sight. Therefore, the explanation of the observed SF slopes for our quasar sample is difficult and requires a complex elaboration of theoretical models to comprehensively account for the variability phenomenon.

\begin{figure}[ht!] 
\centering 
\includegraphics[width=0.37\textheight]{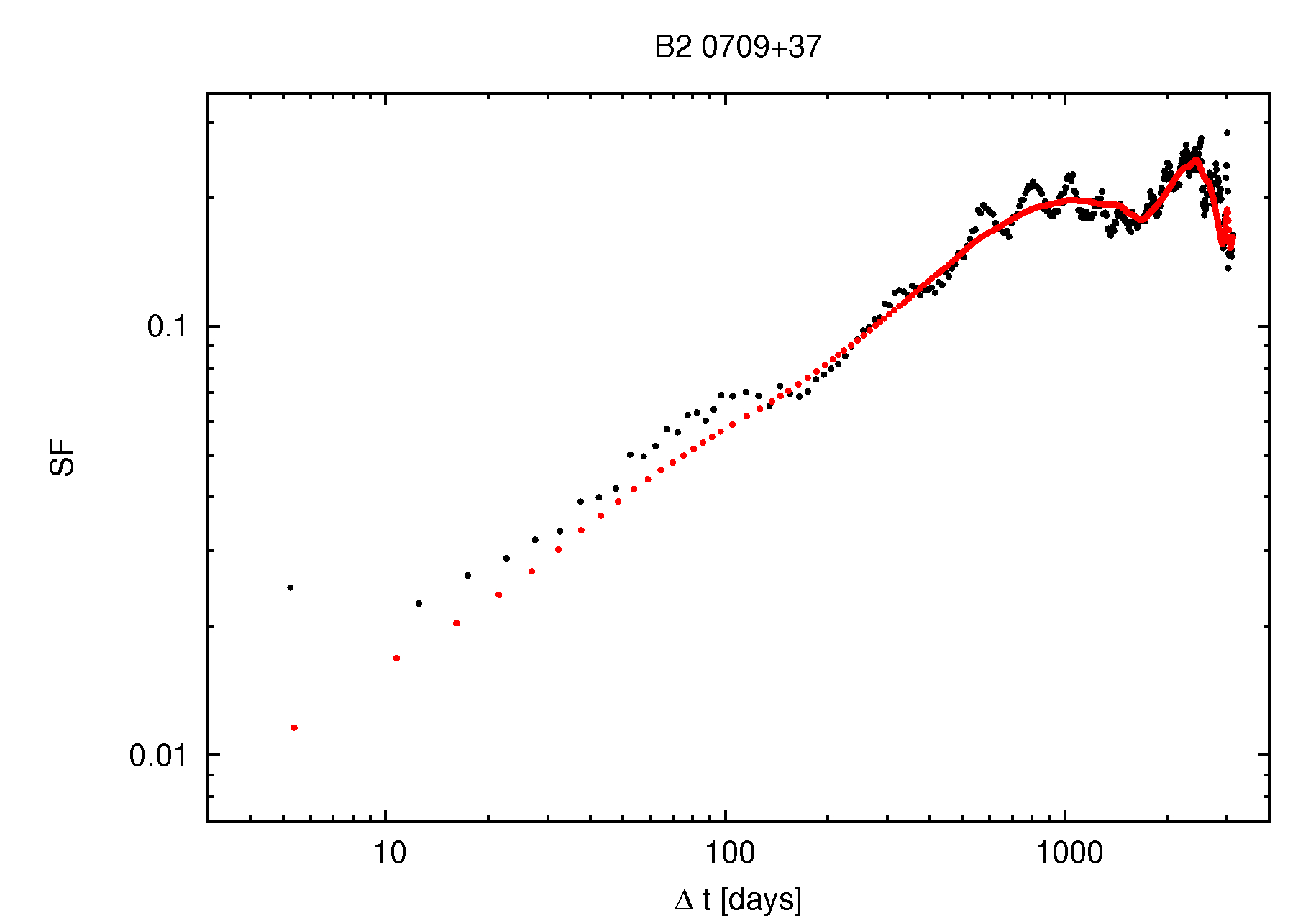}
 
\caption{Structure functions for B2 0709+37. The black points show the SF derived from the original light curve, while the red points show the SF derived from the evenly sampled light curve obtained by interpolation of the observed one.} 
\label{test} 
\end{figure}

As shown by \cite{kozlowski2016} the distribution of SF slopes for AGNs peaks for $\alpha$=0.55$\pm$0.08 with typical decorrelation timescale $\tau$=354 $\pm$ 168 days. Five of the eight quasar light curves from our sample have SF slopes higher than $\alpha$=0.55 and the timescales of variability are also larger in most cases. However, based on 20-year light curves \cite{smith93} found that typical timescales for quasars peaks between 2 and 6 years in a rest frame. It should be noted that to obtain a reliable estimate of characteristic timescales, the length of the data set should be at least a few times longer than the decorrelation timescale \citealt{kozlowski2017}. In the case of our quasar sample, we obtained a sufficiently low $\tau$ compared to the length of the data set for B2 0709+37, [HB89] 1425+267, and [HB89] 1721+343 (the $\tau$ one sixth the length of the data set in a rest frame). 
It was also tested that the slopes or SF breaks may be derived with not sufficient accuracy due to nonregular sampling of the light curve resulting in various types of artefacts in SF shape that can be very misleading in their interpretation \citep{emmanoulopoulos2010}. In particular, such features are visible in the SF of B2 0709$+$37 where the SF has a ``sinusoidal'' shape with peaks repeating every $\sim$230 days. In Figure~\ref{test} we demonstrate how the SF shape of B2 0709$+$37 changes when the data are evenly sampled. For this, the evenly sampled light curve is obtained by linear interpolation between consecutive data points with an interpolation interval of 8 days. The SF calculated for the evenly sampled light curve is smoothed and does not show any artificial peaks. 

\subsection{PSD analysis}\label{sec:psresp}

Power spectral analysis uses Fourier decomposition methods, where the light curve is represented by a combination of sinusoidal signals with random phases and amplitudes, which correspond to various timescales of a source’s variability in the time series (e.g., \citealt{timmer1995}). The observed PSD shows a power-law shape $P(\nu_k)\propto\nu_k^{-\beta}$, where $\beta$ is the slope and $\nu_k$ is the temporal frequency. As the quasar variability can be well described by a stochastic process (e.g. \citealt{kelly2009, kelly2011, kozlowski2016}) the value of $\beta$ represents the type of noise color. The $\beta$=0 is an uncorrelated white-type noise, $\beta$ $\sim$ 1 is a flicker/pink-type noise, and $\beta$ $\sim$ 2 is a damped random-walk red-type noise. The PSD slope is related to the SF slope $\alpha$ as $\beta=4\,\alpha$ \citep[][]{kozlowski2016}, and can be used to discriminate between different variability models.

We use the power spectral response method of \citet{uttley2002} that is widely used in the community \citep{chatterjee2008, maxmoerbeck2014} to derive the PSD slopes for our quasar light curves. We have previously used this method in \citet{goyal2020, goyal2021} and \citet{goyal2022}, so we direct the reader to them for details while we briefly outline the methodology here. The discrete Fourier transform (DFT) of a time series gives the Fourier amplitude as a function of the Fourier frequency. The rms-normalized periodogram is given as the squared modulus of its DFT for the evenly sampled light curve $f(t_i)$, observed at discrete times $t_i$ for a total duration $T$, and consists of $N$ data points:

\begin{eqnarray}\label{eq:psd}
P(\nu_k) = \frac{2 \, T}{\mu^2 \, N^2} \, \Bigg\{ \Bigg[ \sum_{i=1}^{N} f(t_i) \, \cos(2\pi\nu_k t_i)  \Bigg]^2 \nonumber \\ 
+ \Bigg[ \sum_{i=1}^{N} f(t_i) \, \sin(2\pi\nu_k t_i)  \Bigg]^2 \, \Bigg \} 
\end{eqnarray}

where $\mu$ is the mean of the light curve and is subtracted from the flux values, $f(t_i)$. The light curve is evenly sampled through linear interpolation using interpolation intervals 5--10 times smaller than the mean (observed) data sampling interval. The effect of red-noise leak is minimized by multiplying the time series by the Hanning window function \citep[][]{maxmoerbeck2014}. Aliasing is constant and is not effective for red-noise-dominated time series \citep{uttley2002}. The DFT is computed for evenly spaced frequencies ranging from the total duration of the light curve, $T$, down to the mean (observed) Nyquist sampling frequency ($\nu_{\rm Nyq}$). The `raw' periodograms, obtained using Eq.~\ref{eq:psd}), provide a noisy estimate of spectral power (as it consists of independently distributed $\chi^2$ variables with two degrees of freedom (DOF); \citealt{papadakis1993}); therefore, we average a number of them to obtain a reliable estimate, referred as `log--binned' periodograms.  A binning factor of 1.6 is used, with the representative frequency taken as the geometric mean of each bin in our analysis. Finally, the `true' power spectrum in the log-log space is obtained from the `log--binned' periodograms by subtracting the expectation value of $\chi^2$ distribution with 2 DOF (=$-$0.25068; \citealt{vaughan2005}). The `noise floor level' due to measurement uncertainty is computed following \citet{isobe2015}:
\begin{equation}
P_{\rm stat} = \frac{2 \, T}{\mu^2 \, N} \, \sigma_{\rm stat}^2 \, ,\label{eq:poi_psd}
\end{equation}
where $\sigma_{stat}^2= \sum_{j=1}^{j=N} \Delta f(t_j)^2 / N$ is the mean variance of the measurement uncertainties for the flux values, $\Delta f\!(t_j)$, in the observed light curve at times $t_j$, with $N$ denoting the number of data points in the observed light curve.

The best-fit PSD slopes are derived using the power spectral response (PSRESP) method given by \citet{uttley2002}. This method has the advantage of further mitigating the deleterious effects of Fourier transformation and, therefore, is the most robust method for estimating the spectral shape. This method is computationally expensive, as it requires Monte Carlo (MC) simulations of many light curves, mimicking the observed data. Here, an (input) PSD model is tested against the observed PSD. The estimate of the best-fit model parameters and their uncertainties is performed by varying the model parameters. To achieve this, light curves are generated with a known underlying power-spectral shape using MC simulations, and DFT is produced in the same manner as the observed data (Eq.~\ref{eq:psd}). We tested a simple power-law (PL) spectral shape with slopes ranging from 0.1--4.0 with a step of 0.1 and fixed normalization. The quality of fit is assessed by computing two quantities, $\chi^2_{\rm obs}$ and $\chi^2_{\rm dist}$ defined as

\begin{equation}
\chi^2_{\rm obs} = \sum_{\nu_{k}=\nu_{min}}^{\nu_{k}=\nu_{max}} \frac{[\overline{ \log_{10}P_{\rm sim}}(\nu_k)-\log_{10}P_{\rm obs}(\nu_k)]^2}{\Delta \overline{\log_{10}P_{\rm sim}}(\nu_k)^2},
\label{chiobs}
\end{equation}
and
\begin{equation}
\chi^2_{\rm dist, i} = \sum_{\nu_{k}=\nu_{min}}^{\nu_{k}=\nu_{max}} \frac{[\overline{ \log_{10}P_{\rm sim}}(\nu_k)-\log_{10}P_{\rm sim,i}(\nu_k)]^2}{\Delta \overline{\log_{10}P_{\rm sim}}(\nu_k)^2}.
\label{chidist}
\end{equation}
Here, $\log_{10}P_{\rm obs}$ and $\log_{10}P_{\rm {sim, i}}$ are the observed and simulated log-binned periodograms, respectively; $\overline{ \log_{10}P_{\rm sim}}$ and $\Delta \overline{\log_{10}P_{\rm sim}}$ are the mean and standard deviation obtained by averaging a large number of PSDs; $k$ represents the number of frequencies in the Log-binned power spectrum (ranging from $\nu_{k,min}$ to $\nu_{k,max}$), while $i$ runs over the number of simulated light curves for a given $\beta$.

\begin{figure*}[ht!]

\hbox{
\includegraphics[width=0.25\textwidth]{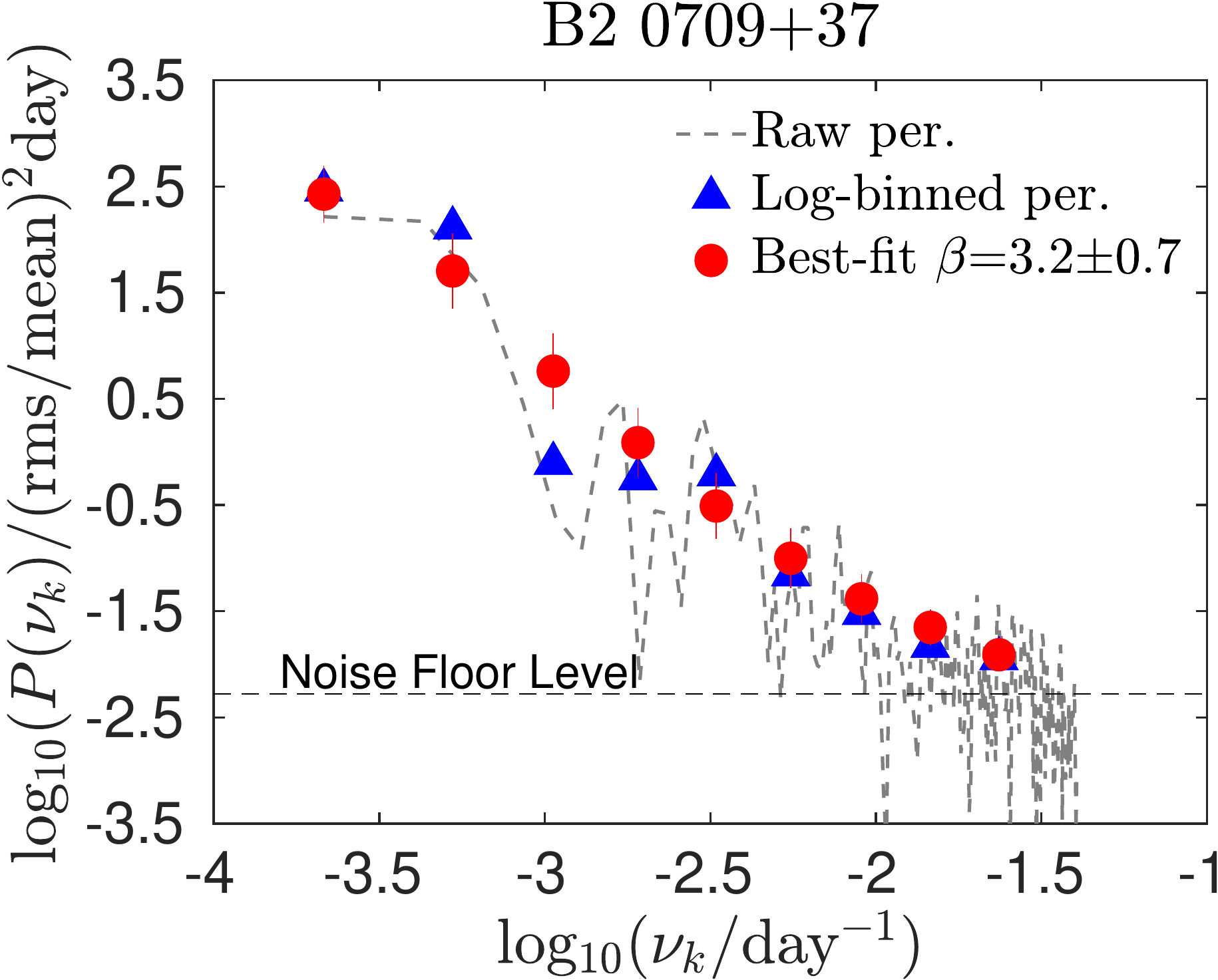}
\includegraphics[width=0.25\textwidth]{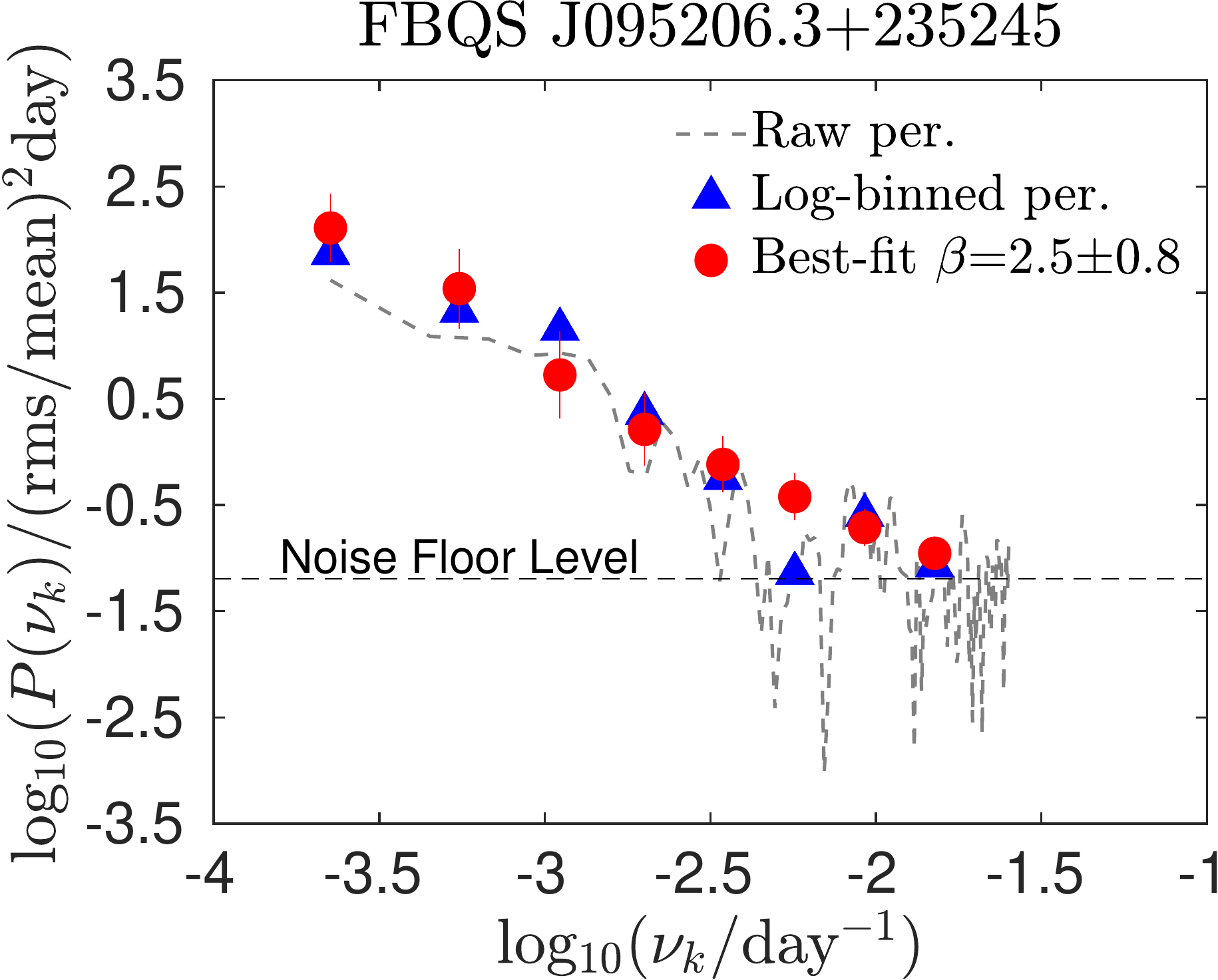}
\includegraphics[width=0.25\textwidth]{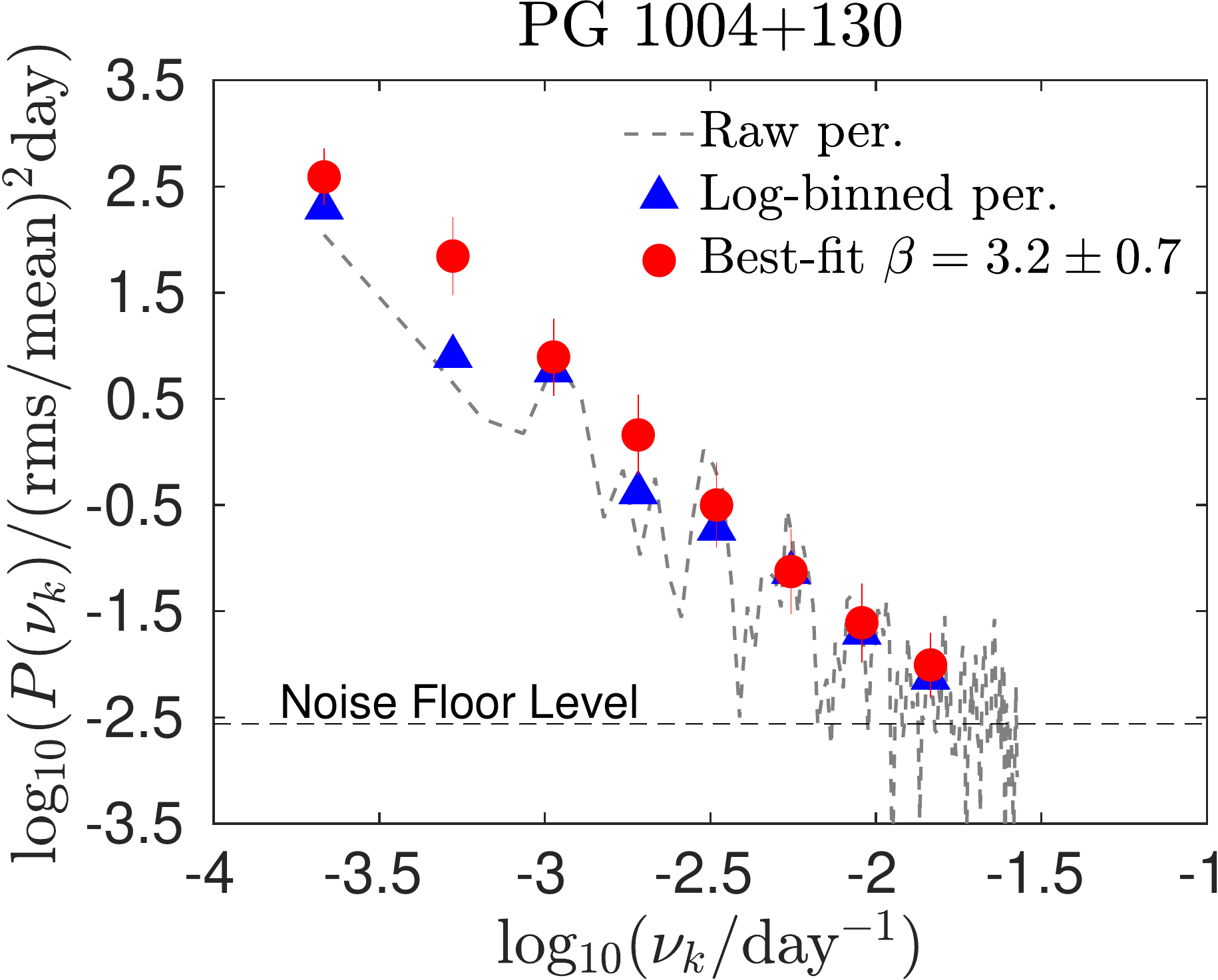}
\includegraphics[width=0.25\textwidth]{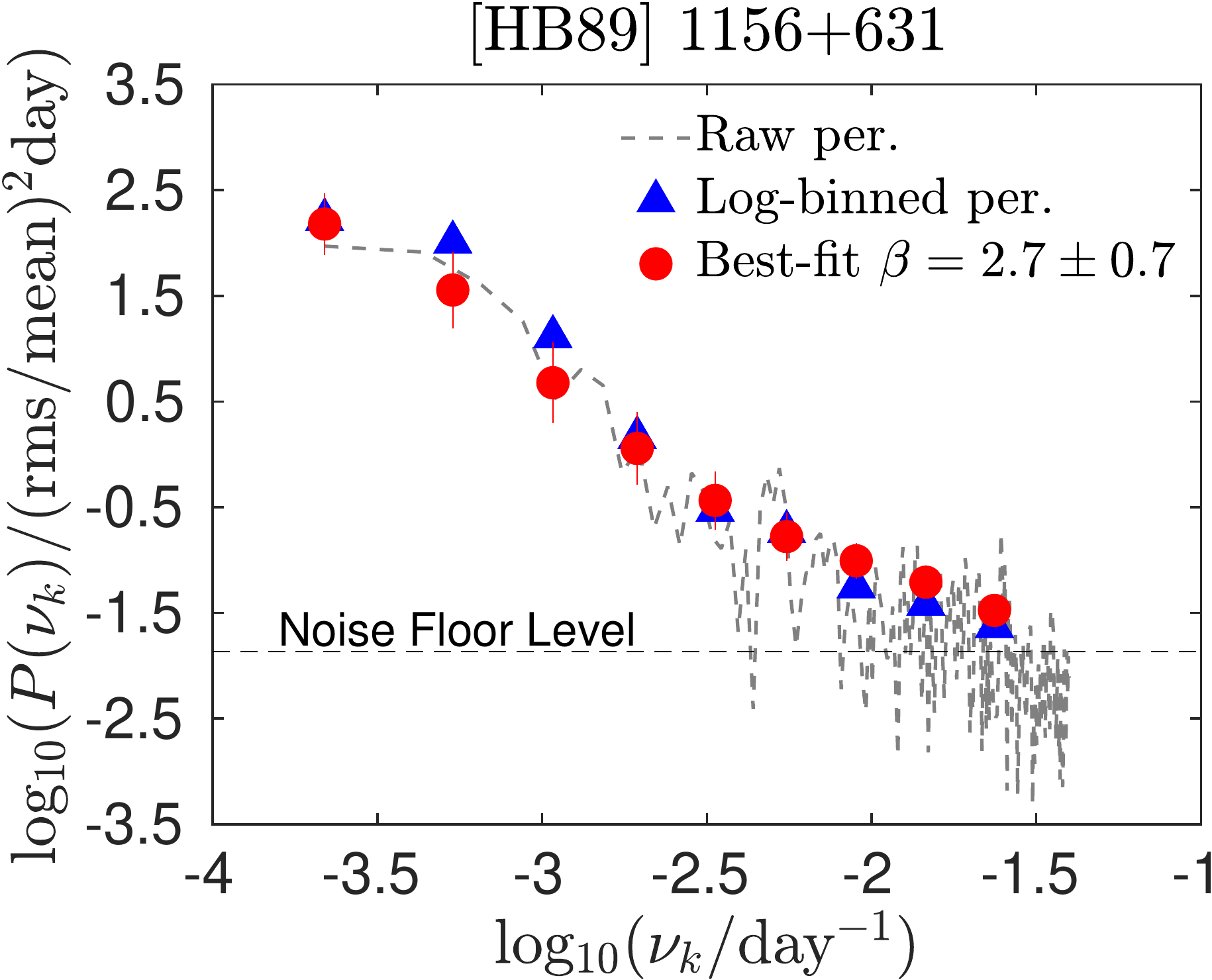}
}
\hbox{
\includegraphics[width=0.25\textwidth]{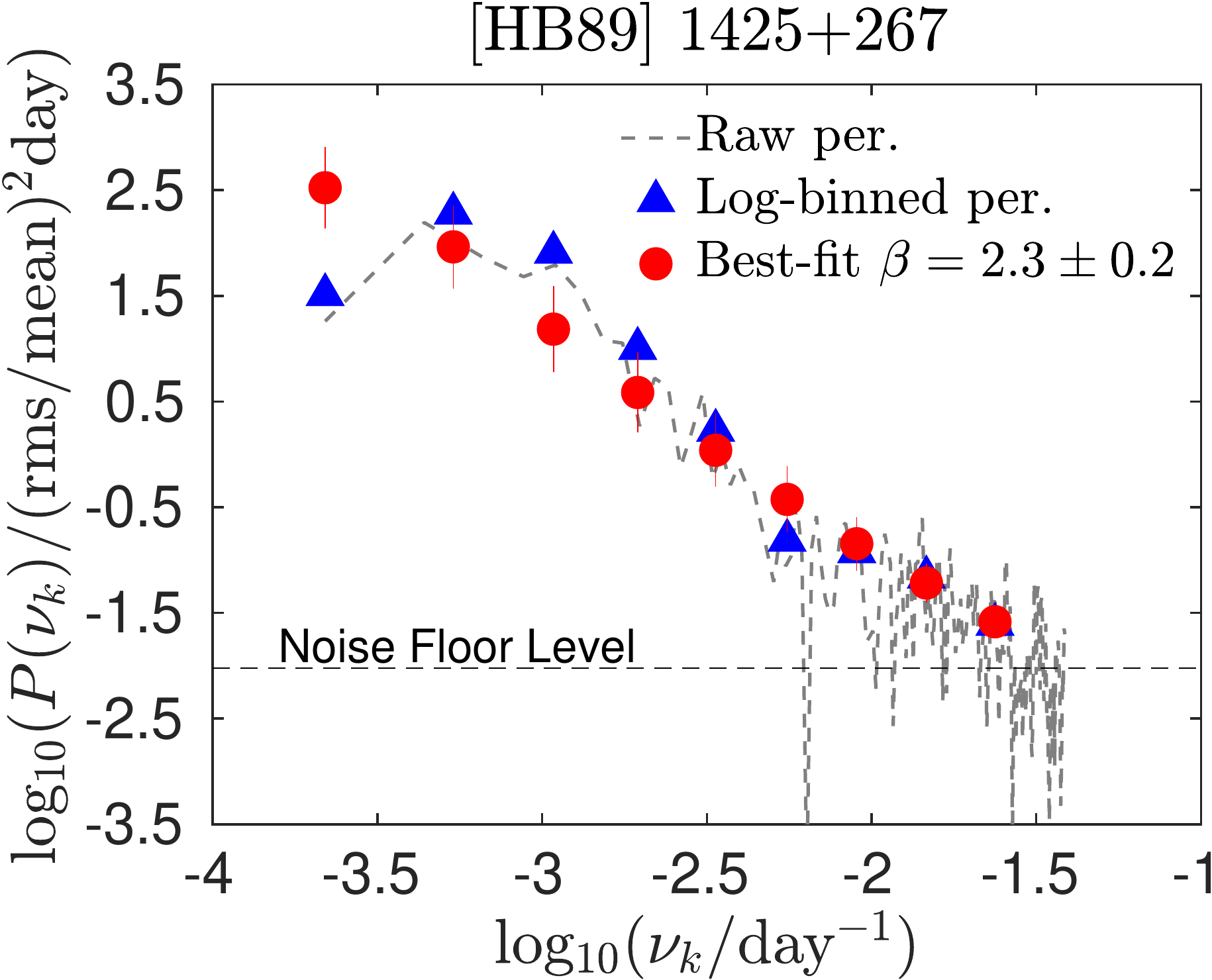}
\includegraphics[width=0.25\textwidth]{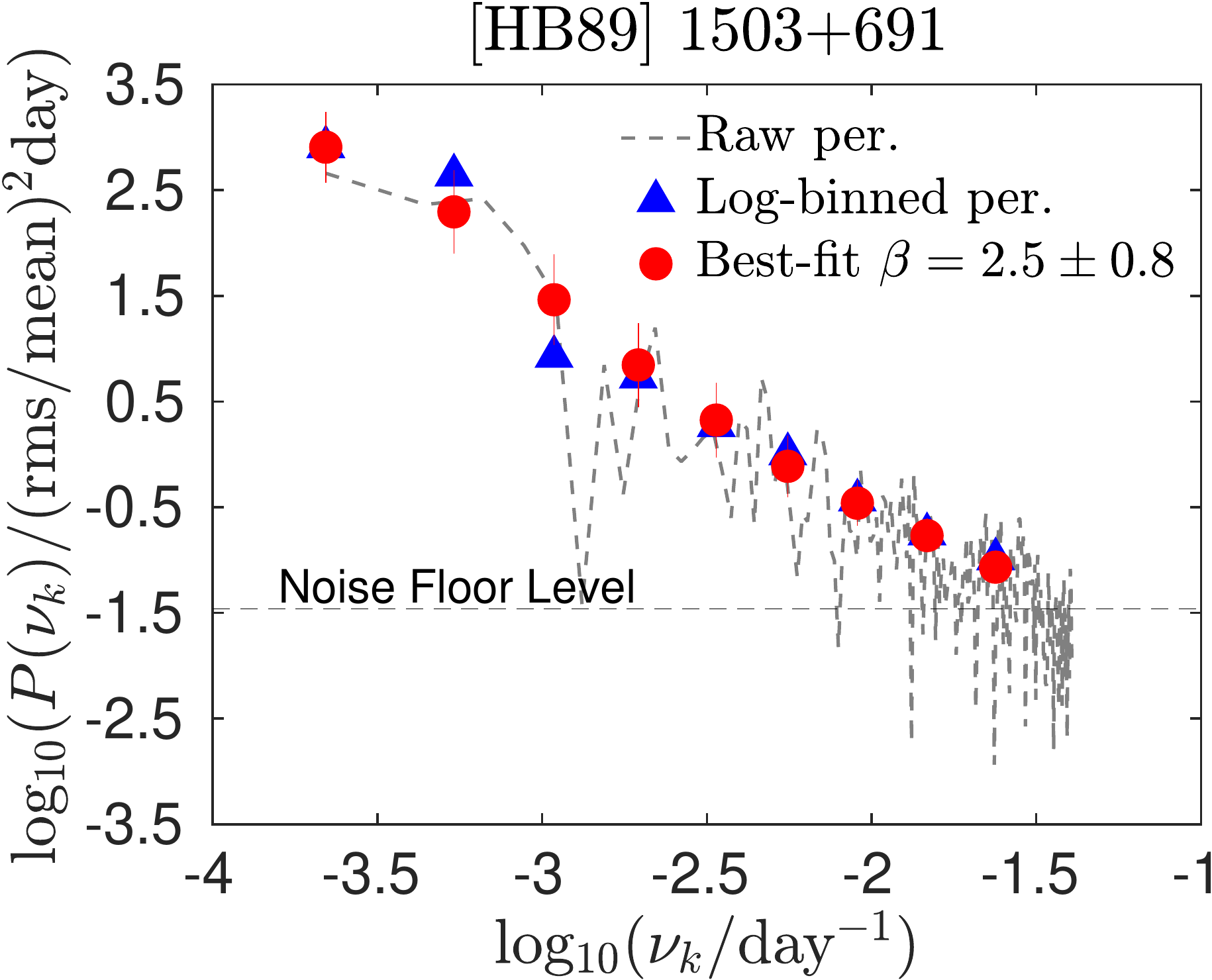}
\includegraphics[width=0.25\textwidth]{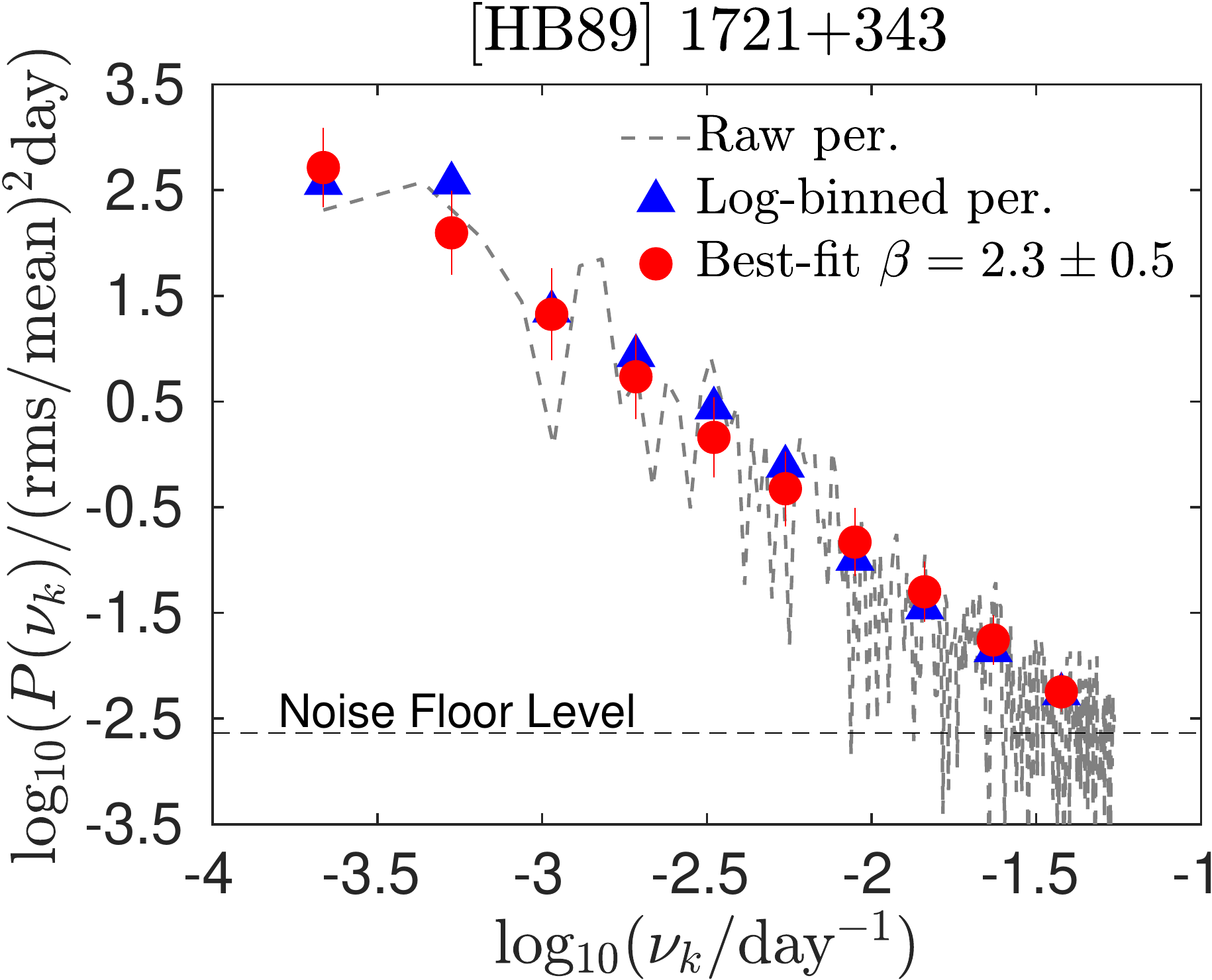}
\includegraphics[width=0.25\textwidth]{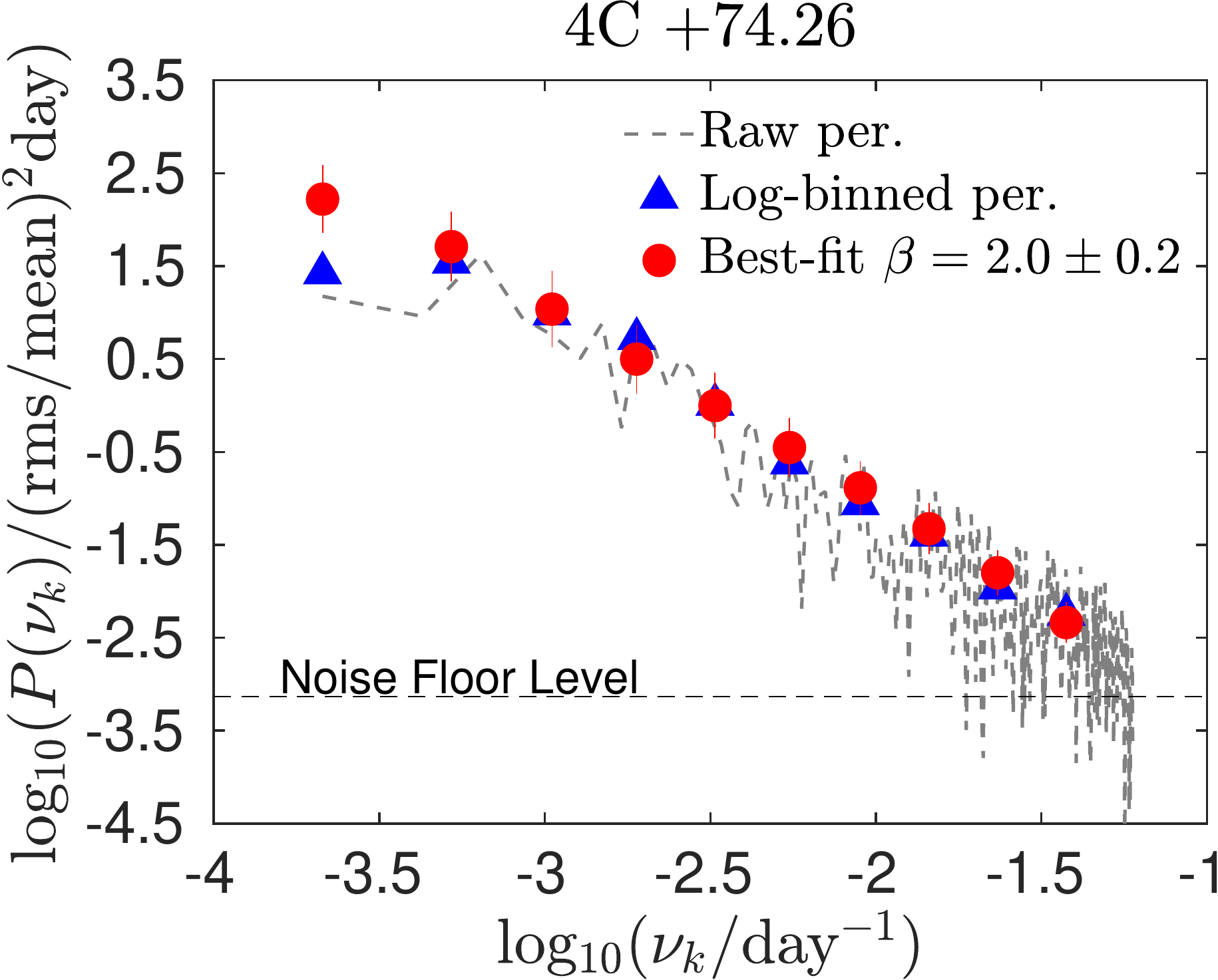}
}
\hbox{
\includegraphics[width=0.25\textwidth]{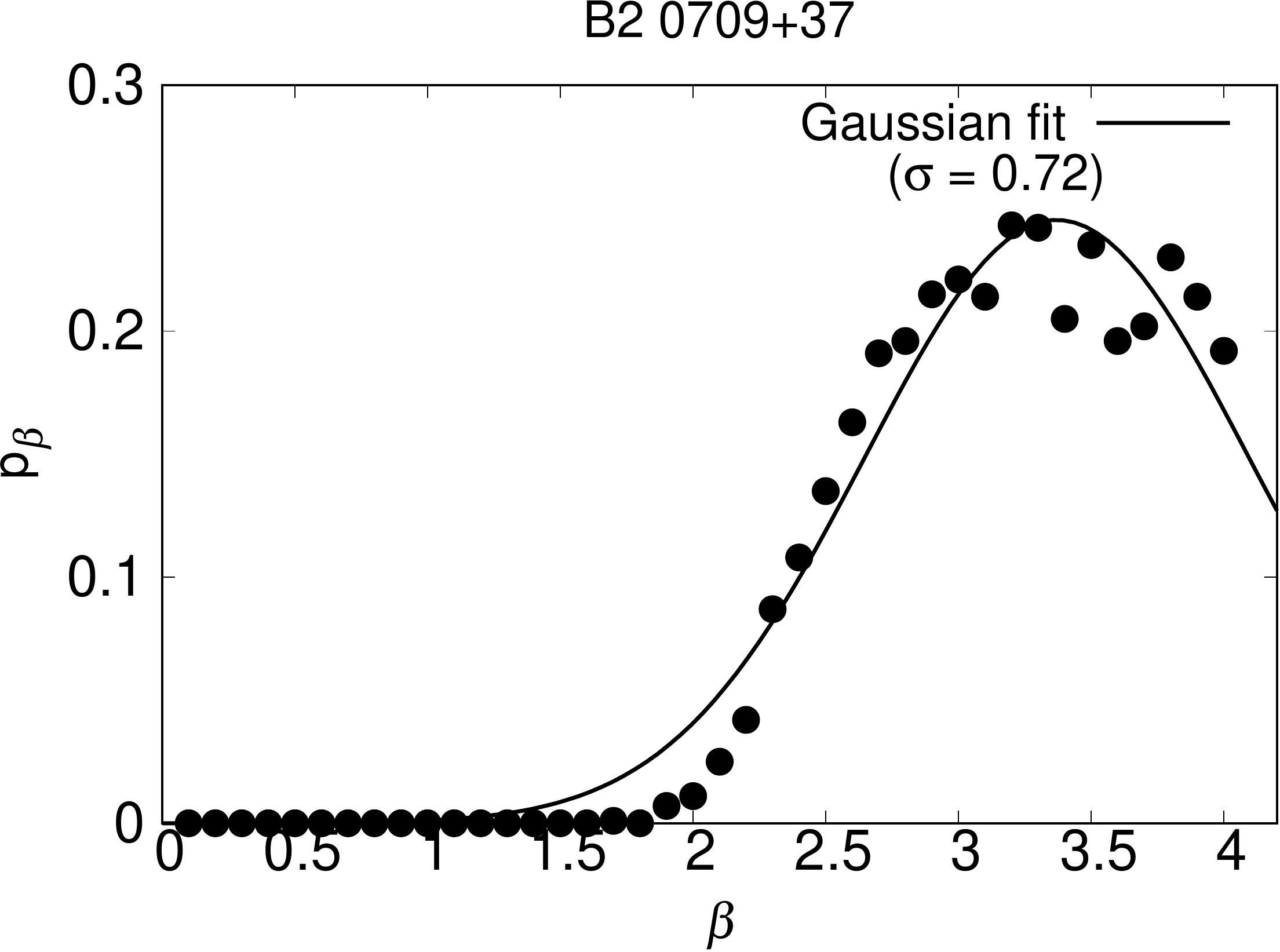}
\includegraphics[width=0.25\textwidth]{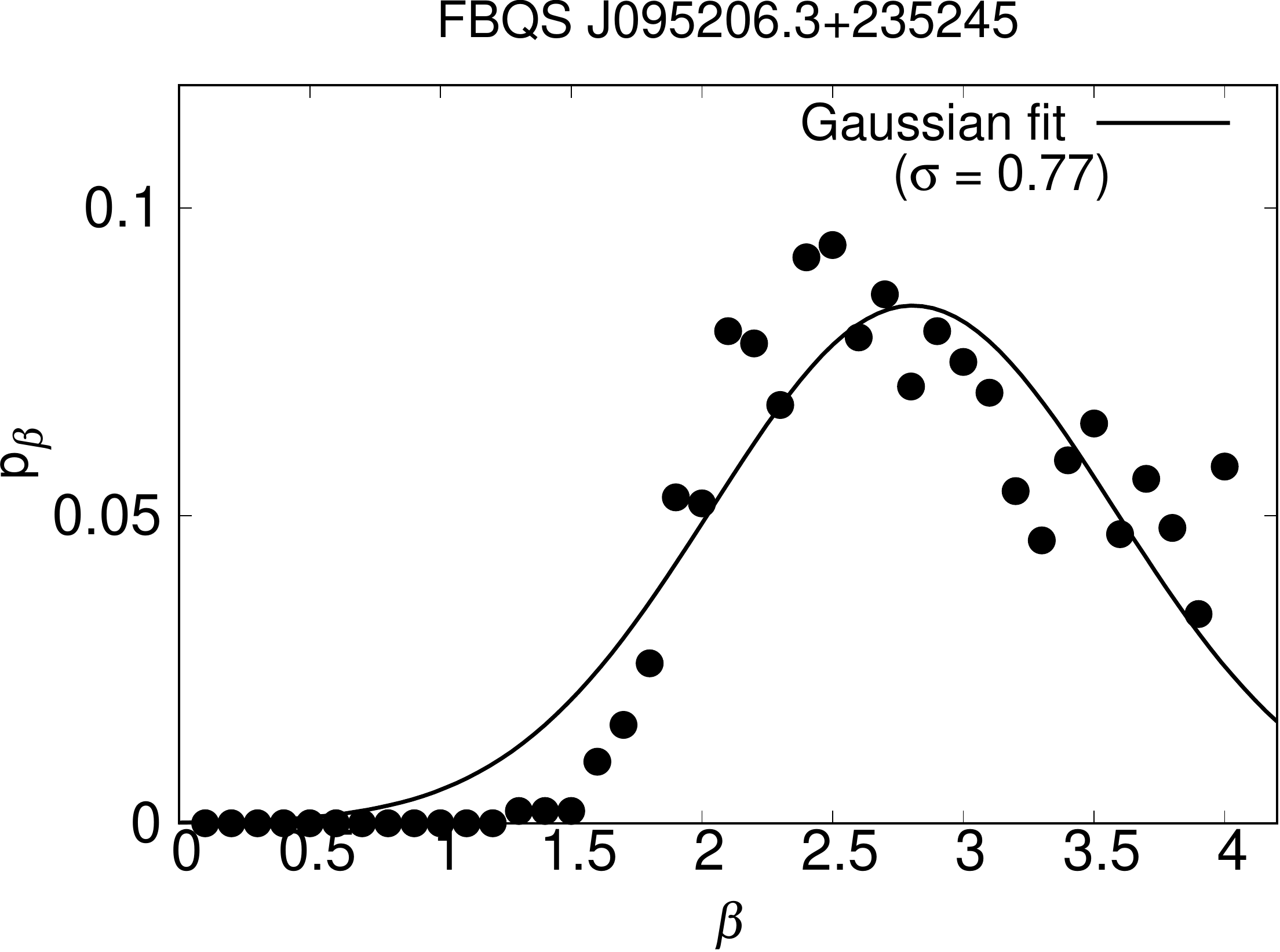}
\includegraphics[width=0.25\textwidth]{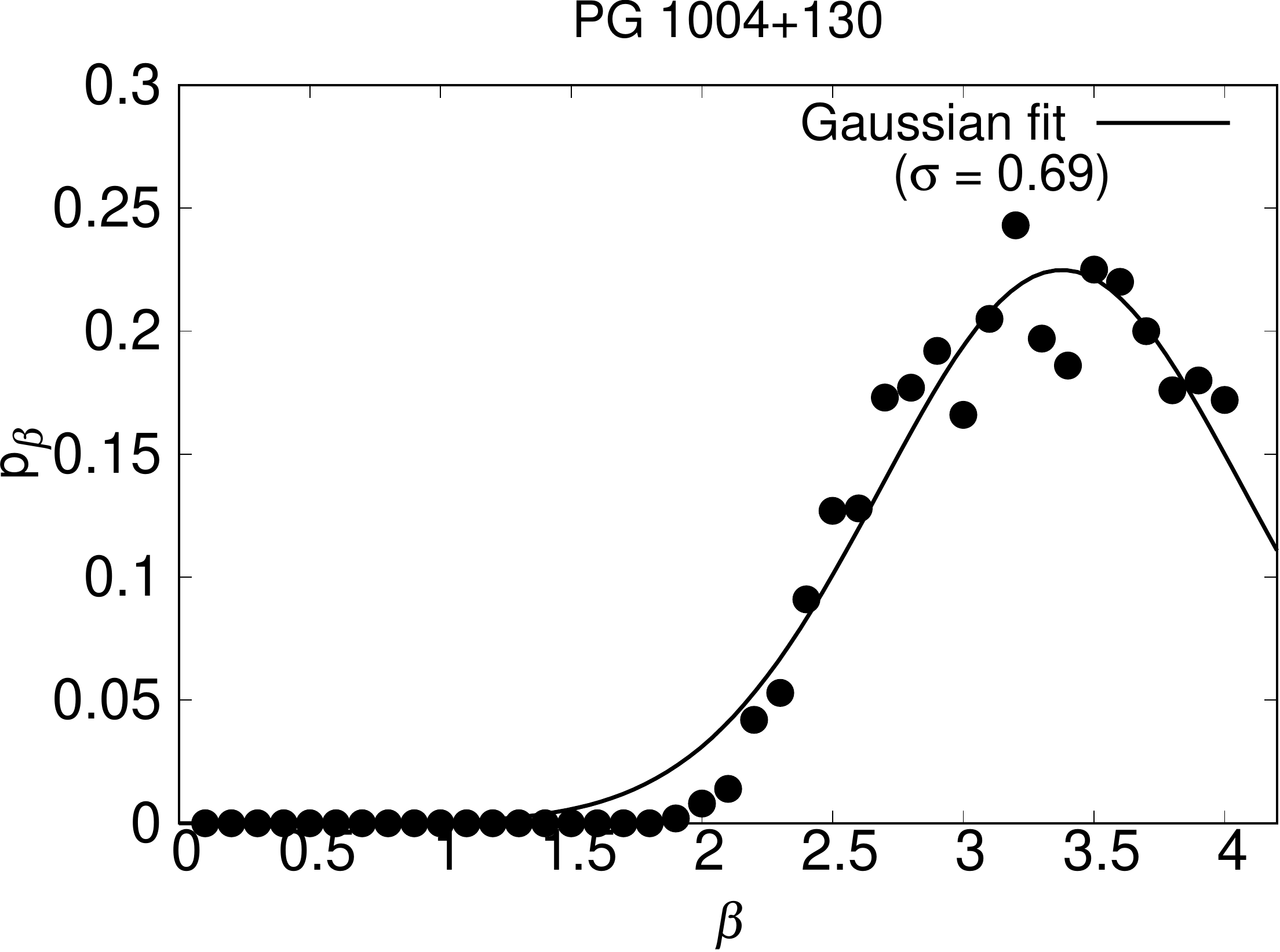}
\includegraphics[width=0.25\textwidth]{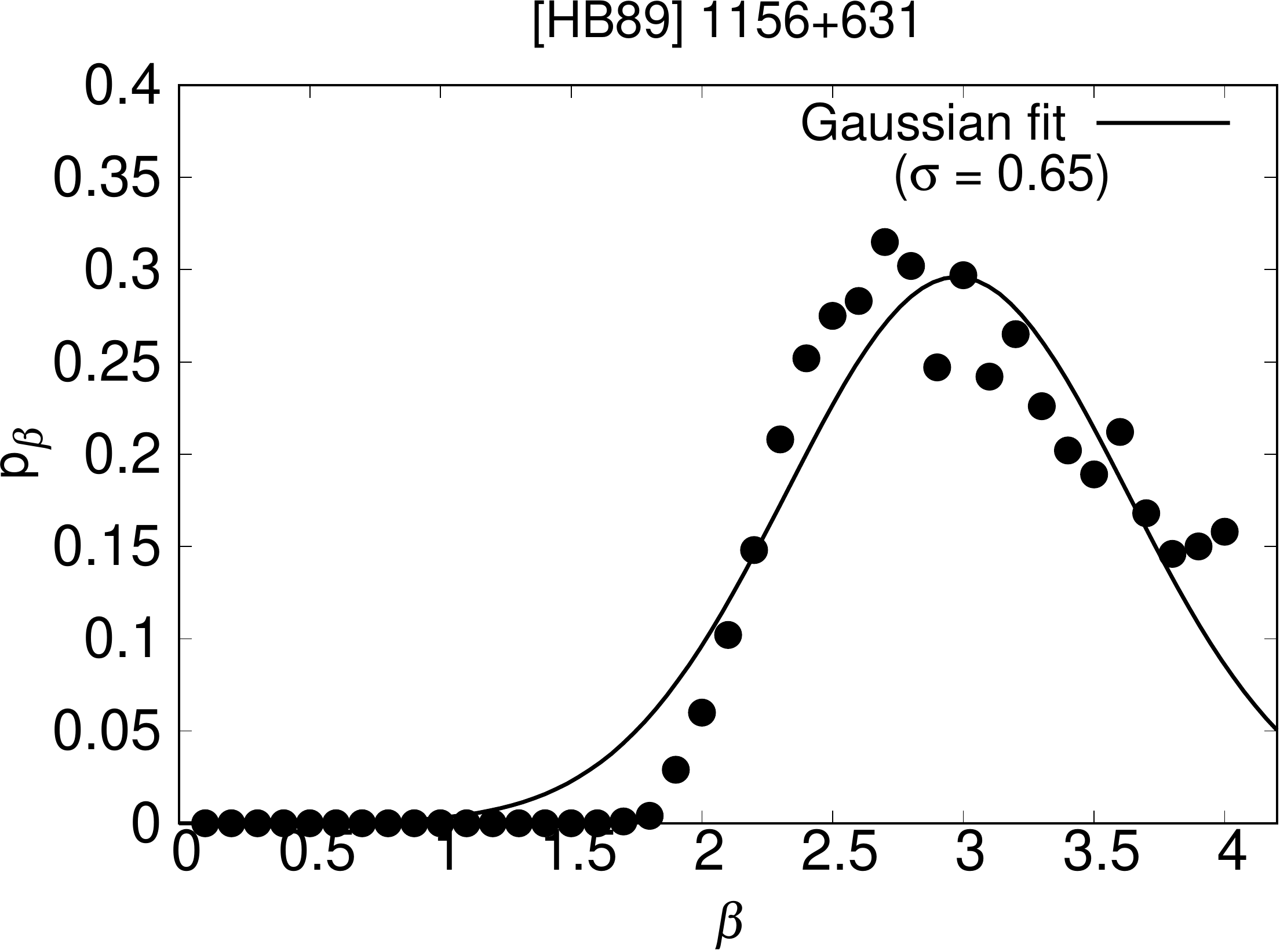}
}
\hbox{
\includegraphics[width=0.25\textwidth]{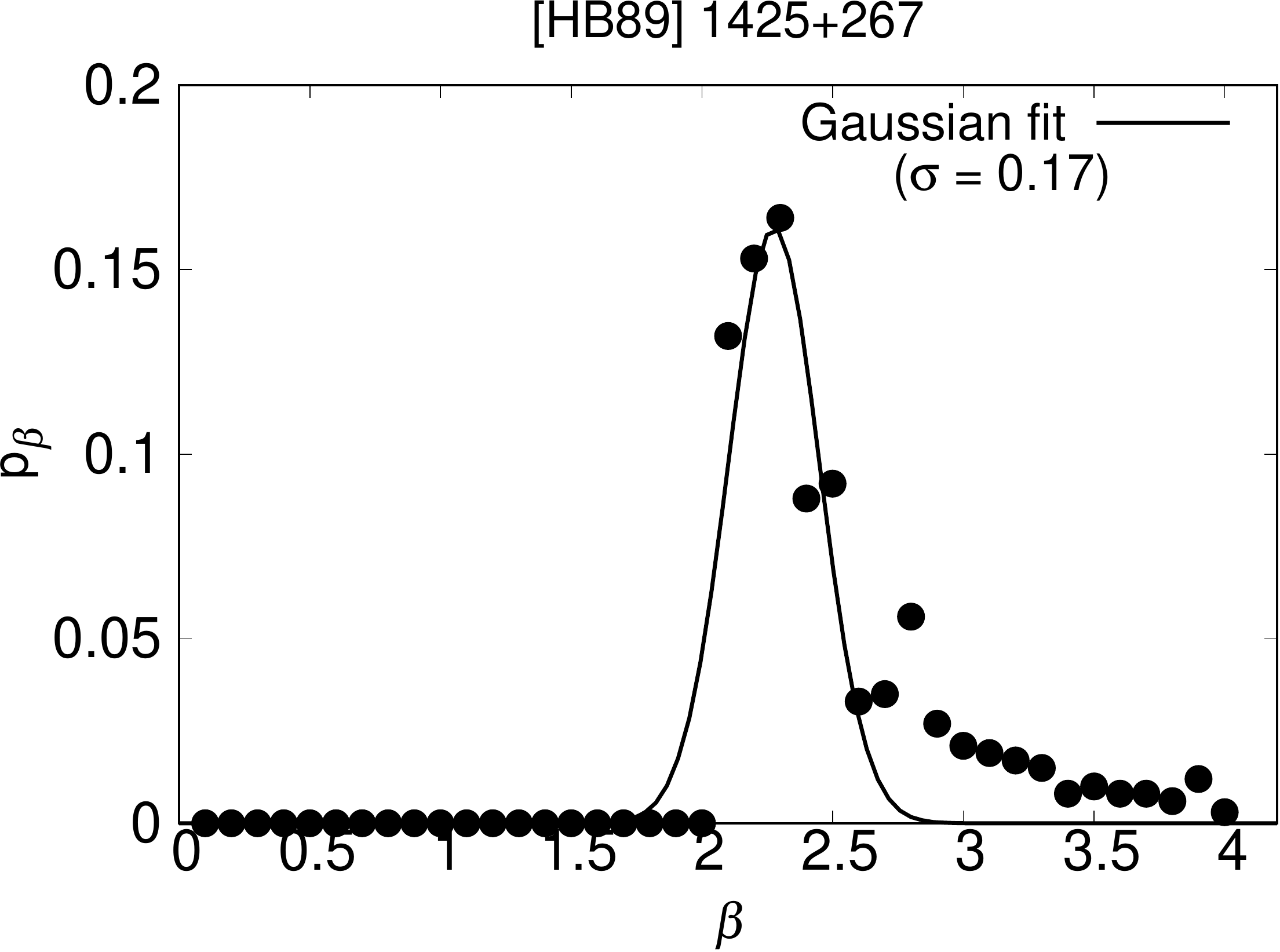}
\includegraphics[width=0.25\textwidth]{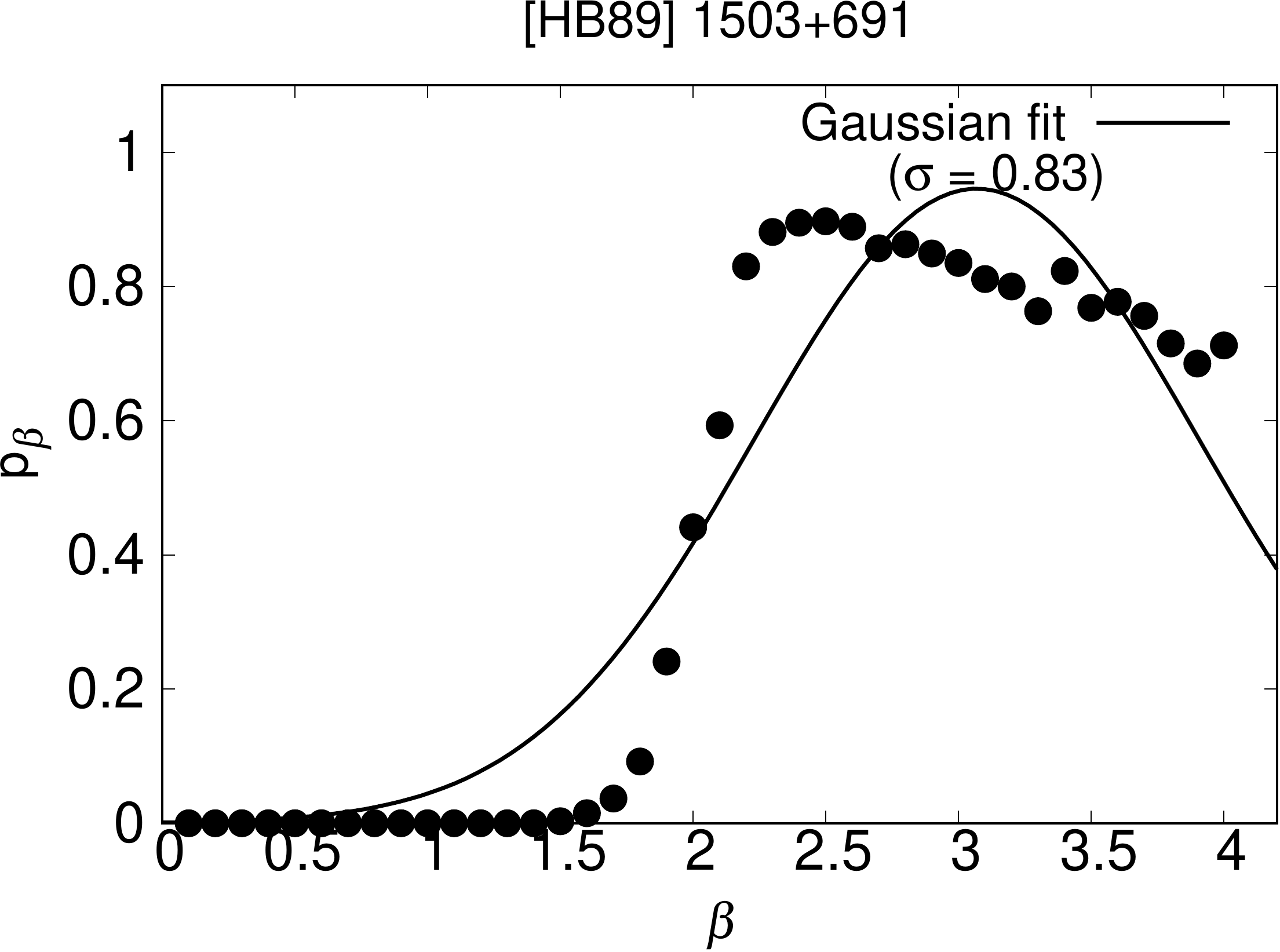}
\includegraphics[width=0.25\textwidth]{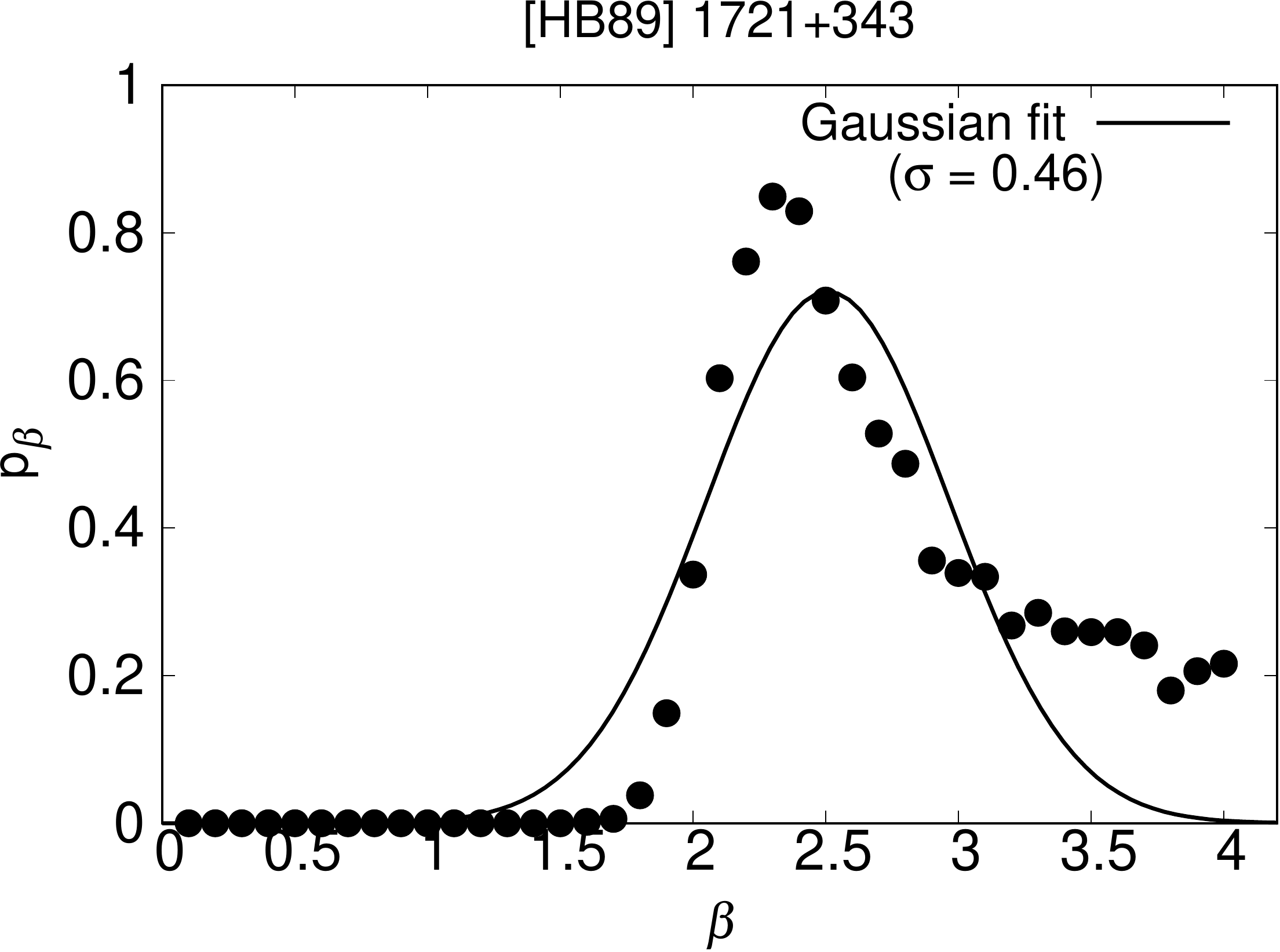}
\includegraphics[width=0.25\textwidth]{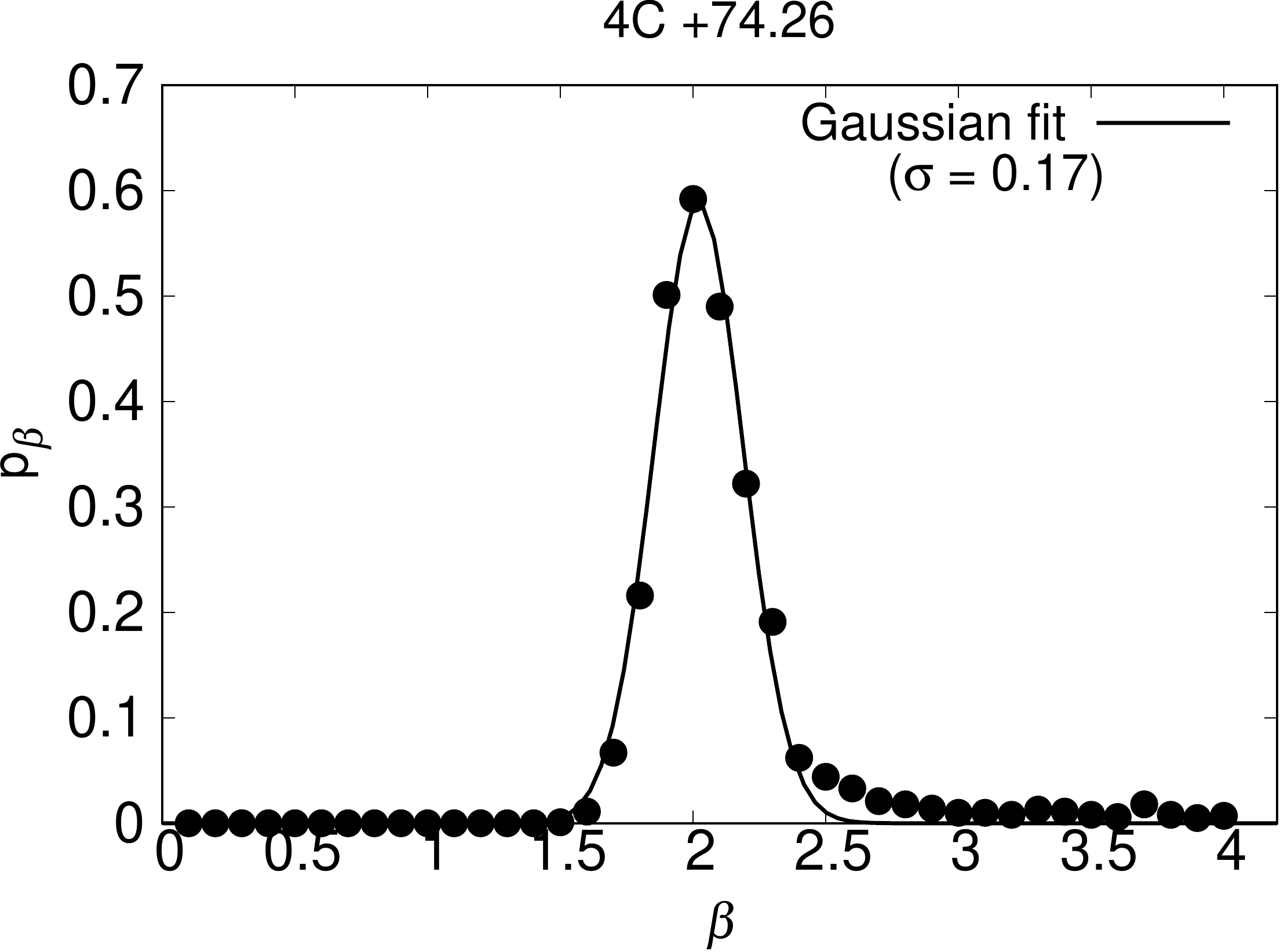}
}
\caption{
Best-fit PSDs and the corresponding $p_\beta$ curves for our quasar light curves. The dashed line shows the raw periodogram while the blue triangles and red circles give the logarithmically binned power spectrum and the best-fit power spectrum, respectively. The error on the best-fit PSD slope corresponds to a 68\% confidence limit, derived from fitting a Gaussian function to the $p_\beta$ curve and is given as standard deviation, $\sigma$. The dashed horizontal line corresponds to the statistical noise floor level due to measurement noise.
}
\label{fig:psd}
\end{figure*}

For a given $\beta$, $\chi^2_{\rm obs}$ is the measure of the offset of the average (input) spectral shape with the observed PSD shape, while $\chi^2_{\rm dist}$ refers to the offset when the observed data are replaced with simulations. A probability, $p_{\beta}$, is given by the percentile of the $\chi^2_{\rm dist}$ distribution above which $\chi^2_{\rm dist}$ is found to be greater than $\chi^2_{\rm obs}$ for a given $\beta$. A large value of $p_{\beta}$ represents a good fit in the sense that a large fraction of random realizations of the model (input) power spectrum can recover the intrinsic PSD shape. Therefore, this analysis essentially uses the MC approach towards a frequentist estimation of the quality of the model compared to the data. For the simulation of quasar light curves, we use the method of \citet{emmanoulopoulos2013} which preserves both the probability density function (PDF) of the flux distribution and the underlying power spectral shape and not just the power spectral shape \citep[e.g. ][]{timmer1995}. For quasar light curve simulations, we used a log-normal PDF of the flux distribution \citep[][]{kelly2009}. Since the light curves are given in units of differential magnitude (Figure~\ref{lc1}), we converted them to fluxes using the arbitrary flux scale, following \citet[][]{goyal2021} for the PSD analysis.   
The errors in the best-fit PSD slope are obtained by fitting a Gaussian function to the probability distribution curves and are given as the standard deviation, $\sigma$, of the Gaussian function. The resulting PSD and the corresponding probability distribution curves for the quasar light curves are plotted in Figure~\ref{fig:psd} while Table~\ref{tab:psd} lists the results of the PSRESP method. The observed PSDs are well-fitted to the single PL function with high confidence ($p_\beta$ $\geq$0.1, except for the quasar FBQS J095206.3$+$235245 for which $p_\beta$= 0.094) and have slopes $\sim$2, over the range of spectral frequency between -3.67 and -1.42\,day$^{-1}$, indicating a red-noise character of variability.

\begin{deluxetable*}{ccccccccc}
\tablenum{3}
\tablecaption{Summary of the observations and the PSD analysis.\label{tab:psd}}
\tablewidth{0pt}
\tabletypesize{\small}
\tablehead{
\colhead{Source Name} & \colhead{Duration} & \colhead{$T_{\rm obs}$} & \colhead{$T_{\rm mean}$} & \colhead{$T_{in}$} &
\colhead{$\rm \log_{10}(P_{stat})$} & \colhead{$\log_{10}(\nu_k) $ range} & \colhead{{$\beta \pm err$}} & \colhead{$p_\beta$$^{\ast}$}\\
 \colhead{} &  \colhead{(Start -- End)} & \colhead{(yr)} &   \colhead{(day)} &    \colhead{(day)} & \colhead{($\frac{\mathrm rms}{\mathrm mean})^2$day} & \colhead{(day$^{-1}$)} & 
}
\decimalcolnumbers
\startdata
  B2 0709$+$37  &  2009 Mar 2 -- 2021 Nov 24  &  12.7  & 12.3 & 0.4    & -2.28   &  -3.66 to -1.62   &3.2$\pm$0.7  & 0.243   \\
  FBQS J095206.3$+$235245 & 2009 Mar 14 -- 2021 May 11 & 12.2 & 19.9 & 0.4 & -1.19 & -3.64 to -1.82 & 2.5$\pm$0.8  & 0.094   \\
  PG 1004$+$130 &  2009 Mar 2 -- 2021 Nov 17 & 12.7 & 18.6 & 0.4 & -2.56 & -3.66 to -1.83 & 3.2$\pm$0.7  & 0.243 \\
  $\rm [HB89]$ 1156$+$631 &  2009 Mar 2 -- 2021 Sep 9 & 12.5 & 12.5 & 0.4 & -1.86 & -3.66 to -1.62 & 2.7$\pm$0.7  & 0.315 \\
  $\rm [HB89]$ 1425$+$267 & 2009 Apr 7 -- 2021 Sep 28 & 12.5 & 12.9 & 0.4 & -2.02 & -3.65 to -1.62 & 2.3$\pm$0.2  & 0.164 \\
  $\rm [HB89]$ 1503$+$691 & 2009 Apr 7 -- 2021 Sep 8 & 12.4 & 12.3 & 0.4 & -1.46 & -3.65 to -1.62 & 2.5$\pm$0.8  & 0.897 \\
  $\rm [HB89]$ 1721$+$343 & 2009 Apr 7 -- 2021 Nov 24 & 12.6 & 9.1 & 0.4 & -2.63 & -3.64 to -1.42 & 2.3$\pm$0.5  & 0.849 \\
  4C $+$74.26 & 2009 Jan 19 -- 2021 Nov 24 & 12.6 & 8.3 & 0.4 & -3.13 & -3.67 to -1.42 & 2.0$\pm$0.2  & 0.592 \\
\enddata
\tablecomments{
(1) name of the quasar,
(2) duration of monitoring (start date -- end date), 
(3) total length of the observed light curve,
(4) the mean sampling interval for the observed light curve (light curve duration/number of data points),
(5) interpolation interval,
(6) the noise level in PSD due to the measurement uncertainty,
(7) the temporal frequency range covered by the  binned logarithmic power spectra,
(8) the best-fit power-law slope of the PSD along with the corresponding errors representing 68\% confidence limit (see Section~\ref{sec:psresp});
(9) the corresponding $p_\beta$. $^\ast$ power law model is considered a bad-fit if $p_\beta$ $\leq$ 0.1 as the corresponding rejection confidence for the model is $\geq$90\% (Section~\ref{sec:psresp}).      
}
\end{deluxetable*}

\section{Discussion}\label{sec:discussion}

\subsection{Structure function vs. PSD analysis}\label{sec:comparison}

The results obtained using the SF and PSD methods for our light curves are notable in two aspects: (1) the slope of the SF curve is related to the PSD curve by $\beta=4\,\alpha$ within the uncertainties reported for each target (Tables~\ref{tab:sf} and ~\ref{tab:psd}) and (2) the SF curves of quasars B2 0709$+$37, [HB89] 1425$+$267, and [HB89] 1721$+$343 show bends (or a plateau) around $\sim$1\,year timescale, indicating a presence of decorrelation timescales (Table~\ref{tab:sf}) while the PSDs of the same sources show a good fit to the single PL form over the entire spectral frequency range covered with no sign of bending at lower frequencies (Table~\ref{tab:psd}). The first result is trivial, as such a dependence between the SF slope (which is a measure of the rms) and the PSD slope (which is a measure of squared rms) is expected \citep[][]{bauer2009} The second result, i.e., SF analysis showing a decorrelation timescale in the light curve for a few quasars, while the PSD analysis indicating no such feature, are inconsistent results, although not completely unexpected. As discussed in detail in \citet{emmanoulopoulos2010}, the SF curves often show breaks that depend on the lengths of the dataset and the underlying PSD shapes. Even for featureless PSD shapes (i.e., single PL forms), the SF breaks could be obtained on timescales corresponding to 1/10 to 1/2 of the length of the time series.  For longer data sets, SF breaks are visible on longer timescales. Thus, the SF break may not reflect the intrinsic variability of the source. Furthermore, to obtain a reliable estimate of characteristic timescales, the length of the data set should be at least a few times longer than the decorrelation timescale \citep{kozlowski2017}. Our comparison of SF and PSD analysis methods also supports the view of \citet[][]{emmanoulopoulos2010} that the results of SF analysis, in particular, the estimation of decorrelation timescales from the finite duration and unevenly sampled time series, should be treated with caution.

\subsection{Multiwavelength light curve analysis of the quasar 4C\,+74.26 and QPO search in optical light curves of our sources}\label{sec:4c74}

 The optical light curve of quasar 4C +74.26 for a duration 2009--2016 from our monitoring program was previously analyzed by \citet{bhatta2018} and \citet{zola2019}. Along with the 15\,GHz radio light curve from Ovens Valley Radio Observatory (OVRO) monitoring and the X-ray light curve from the Swift-BAT Hard X-ray Transient Monitor program of roughly similar duration, \citet{bhatta2018} focused on cross-correlation analysis of light curves between different wavebands. Statistically significant correlated variability was found between optical and radio wavebands with a time lag of 250\,days (radio variation leading to optical variations). \citet{bhatta2018} obtained an optical PSD slope of $\beta$=1.6$\pm$0.2 using the Lomb-Scargle periodogram \citep[LSP; ][]{lomb1976, scargle1982} which is consistent with our estimates of $\beta$=2.0$\pm$0.2 within the reported uncertainties.\\
  \citet{zola2019} analyzed the optical light curve of 4C $+$74.26 and the light curves of all quasars from our sample but for the duration 2009-2018 to search for possible Quasi-Periodic Oscilations (QPOs) using the LSP  and the Weighted Wavelet Z-transform methods (\citealt{foster1996}). No statistically significant ($>$99\% confidence level) QPOs were reported in their analysis for any of the quasars, including 4C\,+74.26. We confirm the lack of QPOs in the optical light curves of our sources in longer duration datasets (2009--2021) as all of them show a good fit to the single PL form with high confidence (Table~\ref{tab:psd}).

\subsection{Projected linear size -- PSD slope correlation}
\begin{figure}[ht!] 
\centering 
\includegraphics[width=0.35\textheight]{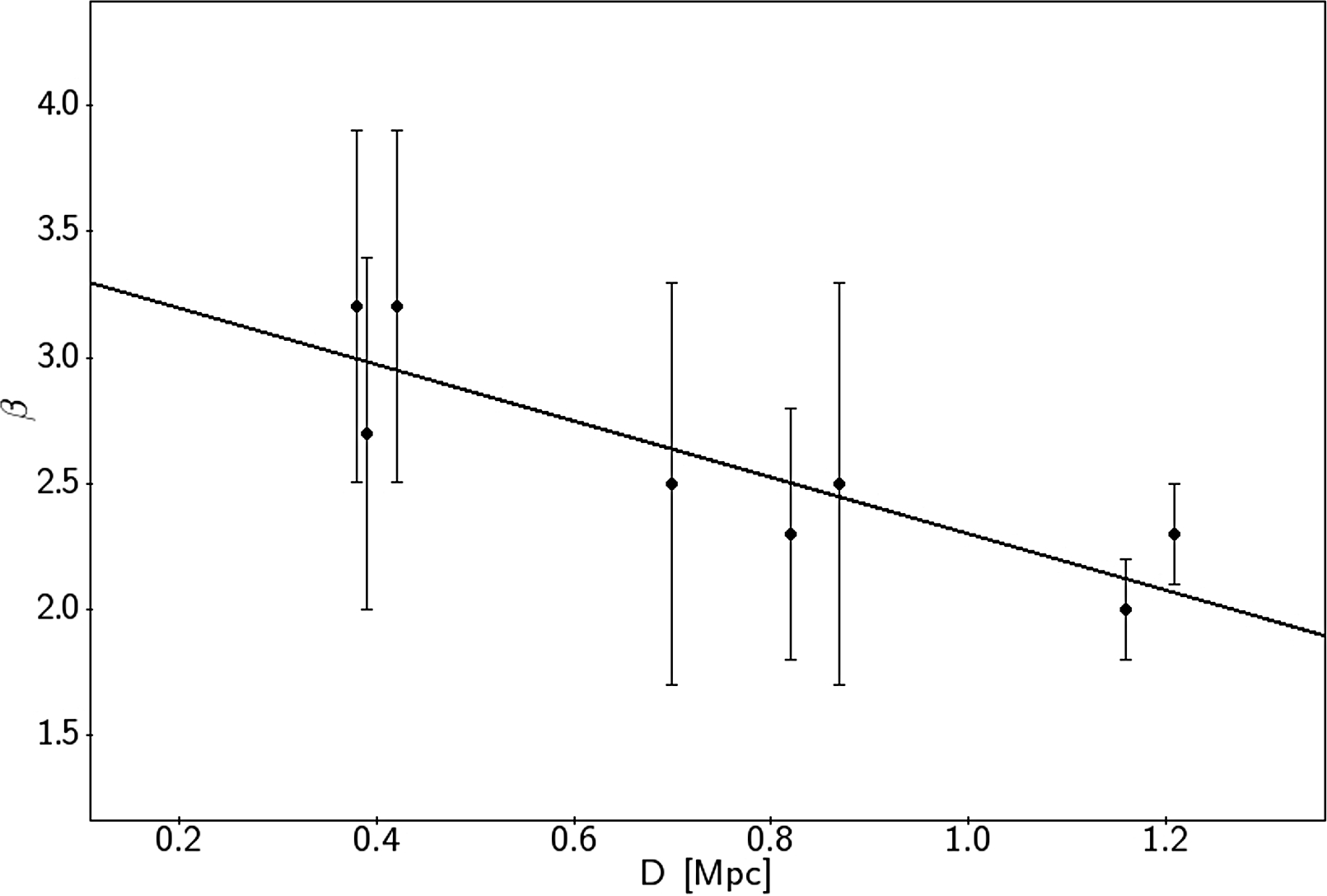}
 
\caption{Anticorrelation between the projected linear size of radio structures (D) and PSD slope ($\beta$).} 
\label{D_b} 
\end{figure}

For the studied quasars, we obtained a significant anticorrelation (correlation coefficient equal to 0.86) between the projected linear size of the radio structure and the PSD slope, i.e., for larger radio sources, the PSD is flatter (see Figure \ref{D_b}). This relationship may indicate that the nature of the variability is related to the size of the radio source, although drawing conclusions based on the results obtained for eight objects is speculation that should be confirmed based on the study of a larger sample of lobe-dominated quasars.

All studied quasars have large radio powers and present large-scale radio structures on the sky plane (Table~\ref{tab:sample}); therefore, it would be interesting to explore the relationship between the optical variability properties (PSD slope) and the linear sizes (Figure~\ref{D_b}).  To check this, we applied Spearman's rank correlation test, which measures the statistical dependence ($p$-value) and the strength of the correlation ($\rho$-value) between the two variables \citep[][]{spearman1904}. A null hypothesis of no correlation is tested against the alternate hypothesis of non-zero correlation at a certain significance level (=0.003, adopted by us).
We obtained a $p$=0.006 and a $\rho$-value=-0.86. Our results indicate that the correlation is significant at a 95--99\% confidence level and the two variables are anti-correlated. This correlation hints that the nature of the optical variability is related to the size of the radio source, although drawing conclusions based on the results obtained for eight objects is speculation that should be confirmed based on the study of a larger sample of lobe-dominated quasars.

\subsection{Physical origin of quasar variability}\label{sec:qsovar}

The optical variations from quasar sources result from instabilities in the accretion disk. In such a case, one expects some characteristic/relaxation timescales beyond which the flux variations should become uncorrelated. Such characteristic/relaxation timescale will appear as flattening in the SF and PSDs curves leading to change of slope from $\geq$1 to $\simeq$0. The natural timescales for a disk are the light-crossing, dynamical, and thermal/viscous timescales which correspond to $\sim$1\,day, $\sim$104\,days and 4.6\,yrs for a standard Shakura and Sunyaev disk (\citealt{shakura1976}) with assumed Eddington ratios of 0.01-0.1, viscosity parameter 0.01, and black hole masses of $10^8$M$_{\odot}$ \citep[][]{kelly2009}. Given the black hole mass for radio quasars typical range of  10$^8$--10$^9$\,M$_{\odot}$ (\citealt{kuzmicz2012}), we expect the SF and PSDs to flatten around a few years timescale, if the fluctuations are driven by thermal instabilities in the accretion disk. Our PSDs cover timescales between $\sim$13\,yr and 2 weeks ($\sim$2--3 times longer than the expected characteristic timescale for a 10$^8$\,M$_{\odot}$ SMBH) and are well-fitted with single PL forms, without any sign of bending on longer timescales.\\
\citet[][]{kelly2009} analyzed the 5--6 year-long light curves of 100 MACHO quasars with SMBH masses in the range 10$^6$--10$^{10}$\,$M_\odot$ using the stationary Ornstein-Uhlenbeck (OU) process or a random-walk type noise process which give a well-defined `relaxation time' of the process driving the variability.  On timescales longer and shorter than the relaxation time, the resulting PSDs exhibit slopes equal to 0 and 2, respectively. In their analysis, relaxation timescales of $\sim$100--1000\,days were obtained. It was interpreted that stochastic fluctuations in magnetic field intensity are dissipated in the disk, transferring energy from the magnetic field to heat in the plasma and creating thermal fluctuations in the accretion disk. This leads to stochastic fluctuations in the luminosity; however, the disk cannot react to heat content changes on timescales shorter than the thermal timescales and leads to damped (or red-noise) type variability. The disk only forgets the input heat content on timescales longer than the thermal timescales, thereby creating uncorrelated (or white-noise) type variability. Although the optical variability of our quasars is also characterized by random-walk type processes ($\beta\geq$2; Table~\ref{tab:psd}), we do not recover the change of slope in the PSDs, pertaining to thermal timescales in the disks of our sources. This could be due to the fact that timescales longer than $\geq$1,000\,day are poorly sampled in our data which leads to a large scatter in their estimation in our PSD modeling.

\section{Summary and final remarks}

In this study, we present the long-term optical variability light curves of eight lobe-dominated quasars from our monitoring program started in 2009. The observed light curves were obtained with a typical sampling period of a few days and cover a total duration of $\sim$13\,yr, showing magnitude variations in the range of 0.3 and 1 mag and present unique datasets in terms of roughly uniform sampling of light curves and a homogeneous reduction procedure. We characterize the optical variability using the two widely used complementary analysis methods, SF and PSDs. Our main findings are as follows:\\
\begin{enumerate}
    \item The slopes obtained from the SF and PSD analysis correspond well to each other. The statistical character of optical variability is red-noise or damped-random walk type process over weeks to $\geq$decade timescales for the quasar sources.
    
    \item The SF analysis of light curves of a few quasar sources indicates the presence of decorrelation timescales, which is not supported by the corresponding PSD analysis. Our complementary approach to variability analysis highlights the need that such approaches should be commonly employed; as far as the detection of spurious features in the time series is concerned, which can lead to erroneous interpretation of the physical origin of variability.
    
    \item We did not detect QPOs in the analysed light curves of quasar sources.  
    
    \item We obtained a mild hint that PSD slopes are anticorrelated with the linear sizes of the radio structures. Despite the fact that anticorrelation is statistically significant, it was obtained on the basis of 8 points, so in fact, the result may be an effect of sample selection. For this reason, it should be confirmed for a larger sample of objects.
    
    \item The absence of a characteristic/relaxation timescale (i.e., flattening of PSD slope on longer timescales) in the analyzed light curves, now extending up to the $\geq$decade timescale indicates that very likely thermal instabilities are the driver behind the variability; although these relevant timescales cannot be revealed due to large statistical scatter in their estimation because of the small number of long temporal frequencies in the analyzed PSDs.
\end{enumerate}

As all our sources present large-scale radio structures in the sky, it is tempting to draw an analogy with X-ray binaries which also present radio jets in the soft/high state and show QPOs in the low/hard state \citep[see for a review][]{mchardy2006}. If we employ a simple scaling relation of BH mass from X-ray binaries to our quasars (mass $\sim10^8$\,M$_{\odot}$, $L_{\rm bol}\sim\,1\,L_{Edd}$, where $L_{Edd}$ is the Eddington luminosity), we expect the break timescales (alternatively, decorrelation timescales) to be around millions of years. Naturally, such timescales are beyond conventional methods of monitoring. We end with a final remark that to pin-point the physical origin of quasar optical variability, one needs to study light curves covering a few decades timescales (longer than the expected relaxation timescale due to thermal fluctuations) with a sampling interval of 1--4\,days at least. In such cases, one might reveal a break in PSD beyond the expected thermal timescales, which will be possible with the new facilities such as Vera C Rubin Observatory.  

\begin{acknowledgments}
We thank Szymon Koz{\l}owski for valuable comments and discussions.\\
This project was partially supported by the Polish National Science Centre (NCN) under decision UMO-2018/29/B/ST9/01793. A.G. acknowledges support from the NCN through the grant 2018/29/B/ST9/02298. The quasar light curve simulations were performed using the PLGrid Infrastructure under the computing grant plglcsims.
\end{acknowledgments}

\vspace{5mm}
\facilities{Cassegrain:50\,cm, Suhora:60\,cm, SKYNET, RCOP:40\,cm, MDRS:40\,cm, DAO:40\,cm  }

\software{IRAF \citep[][]{tody1986, tody1993}
          }

\bibliography{sample631}{}
\bibliographystyle{aasjournal}

\appendix
\restartappendixnumbering 

\section{Comparison stars}\label{appendix1}

In Table~\ref{tab:tab_comp} we list the coordinates and $r$-band Pan-STARRS magnitudes (\citealt{chambers2016}) for comparison and check stars used for each target in our observations. 

\begin{table*}[htp!] 
\caption{Comparison and check stars used in differential light curves.\label{tab:tab_comp}} 
\begin{tabular}{c|ccc|ccc} 
\hline \hline
Source Name	&  \multicolumn{3}{c|}{Comparison star} & \multicolumn{3}{c}{Check star}\\
\hline 
	& RA(J2000) & Dec(2000)                           &  $r$-mag   & RA(2000) & Dec(2000)                          &  $r$-mag \\ 
	& (h m s) 	 &($\rm^{o}$ $^{\prime}$ $^{\prime\prime}$) & (mag) & (h m s)        &($\rm^{o}$ $^{\prime}$ $^{\prime\prime}$)&  (mag)  \\
\hline
B2 0709$+$37	& 07 12 48.50  &  $+$36 52 49.42  & 13.5243    &  07 13 38.93  &  $+$36 56 19.43  & 13.7965\\
FBQS J095206.3$+$235245	& 09 52 03.61  &  $+$23 53 55.11  & 14.9996    &  09 51 43.60  &  $+$23 55 37.19  & 15.6902\\
PG 1004$+$130	& 10 07 47.64  &  $+$12 46 38.96  & 14.0206    &  10 07 23.19  &  $+$12 44 54.28  & 14.2065\\
$\rm[HB89]$ 1156$+$631	& 11 58 50.39  &  $+$62 53 05.86  & 14.8733    &  11 58 46.30  &  $+$62 51 58.65  & 15.1073\\
$\rm[HB89]$ 1425$+$267	& 14 27 27.92  &  $+$26 29 10.13  & 15.0046    &  14 27 30.06  &  $+$26 36 05.26  & 14.3432\\
$\rm[HB89]$ 1503+691 	& 15 03 10.47  &  $+$68 58 09.50  & 15.6052    &  15 04 30.66  &  $+$68 52 09.56  & 15.0096\\
$\rm[HB89]$ 1721$+$343 	& 17 23 30.52  &  $+$34 18 39.11  & 13.9844    &  17 23 21.33  &  $+$34 15 54.52  & 14.3638\\
4C $+$74.26	& 20 43 09.07  &  $+$75 05 56.27  & 14.5511    &  20 42 55.65  &  $+$75 09 42.53  & 14.9003\\ 
\hline
\end{tabular}
\end{table*} 

\section{Differential light curves}\label{appendix2}
Table with light curves for each quasar is published in their entirety in the machine-readable format. Portion of data is shown in Table \ref{tab:tab_0713} for guidance regarding its form and content. Column description: 
(1) -- Quasar name, (2) -- heliocentric Julian date, (3) -- differential R-band magnitude between QSO (denoted as V) and comparison star (denoted as C); (V-C) (4) -- error of (V-C), (5) -- differential R-band magnitude between comparison and check star (denoted as C1); (C-C1), (6) -- error of (C-C1), (7) -- telescope used for observation: KR50 -- 50 cm telescope of the Astronomical Observatory of the Jagiellonian University; SUH -- 60 cm telescope at the Mt. Suhora Astronomical Observatory; CDK -- 50 cm telescope of the Astronomical Observatory of the Jagiellonian University (Skynet); DSO -- 40 cm telescope of the Dark Sky Observatory (Skynet); RRRT -- 60 cm Rapid Response Robotic Telescope of the Fan Mountain Observatory (Skynet); YERKES -- 60 cm or 100 cm telescope of the Yerkes Observatory (Skynet); NSO -- 40 cm telescope of the Northern Skies Observatory (Skynet); MLC -- 40 cm telescope of the Montana Learning Center (Skynet); RCOP -- 40 cm telescope of the Perth Observatory  (Skynet); MDRS -- 40 cm telescope of the Mars Desert Research Station (Skynet); DAO -- 40 cm telescope of the Dolomiti Astronomical Observatory.

\setcounter{table}{0}
\renewcommand{\thetable}{B\arabic{table}}
\newpage
\begin{longtable}{ccccccc}
\caption{Differential R-band magnitude light curves for studied quasars (first ten lines).\label{tab:tab_0713}}\\\hline\hline
Quasar name & JDHEL &  V-C &  V-C\_er & C-C1 & C-C1\_er& Tel.\\
 & days	& mag	& mag	& mag	& mag	&\\
(1) & (2) & (3) & (4) & (5) & (6) & (7)\\
\hline

B2 0709+37 & 2454893.25582  &  2.329  &  0.027  &  -0.247  &  0.003  &  KR50\\
B2 0709+37 & 2454905.26892  &  2.331  &  0.022  &  -0.239  &  0.007  &  KR50\\
B2 0709+37 & 2454908.36548  &  2.325  &  0.016  &  -0.238  &  0.006  &  KR50\\
B2 0709+37 & 2454909.35456  &  2.343  &  0.020  &  -0.233  &  0.006  &  KR50\\
B2 0709+37 & 2454912.31558  &  2.313  &  0.025  &  -0.233  &  0.006  &  KR50\\
B2 0709+37 & 2454919.27040  &  2.319  &  0.019  &  -0.239  &  0.005  &  KR50\\
B2 0709+37 & 2454928.31970  &  2.319  &  0.044  &  -0.240  &  0.013  &  KR50\\
B2 0709+37 & 2454930.29238  &  2.316  &  0.045  &  -0.218  &  0.013  &  KR50\\
B2 0709+37 & 2454942.30533  &  2.317  &  0.022  &  -0.246  &  0.006  &  KR50\\
B2 0709+37 & 2454943.30372  &  2.317  &  0.018  &  -0.247  &  0.004  &  KR50\\
\hline
\end{longtable}


\end{document}